\documentclass[sigconf]{acmart}

\usepackage{epsfig,epstopdf,endnotes,tabu,longtable}
\usepackage{supertabular,lipsum,verbatim}
\usepackage{float}
\usepackage{tabularx,booktabs,threeparttable,alltt,enumitem}
\usepackage{url}
\usepackage{hyperref}
\hypersetup{breaklinks=true}
\usepackage{balance,adjustbox,amsmath,amsfonts,xfrac}
\usepackage{subfigure,xcolor,alltt,enumitem,array}
\usepackage{fullpage,mathptmx,microtype, lmodern}
\usepackage[ruled,lined]{algorithm2e}
\usepackage[T1]{fontenc}
\usepackage{multirow,array}
\usepackage[disable]{todonotes}

\acmConference[DYNAMICS]{DYnamic and Novel Advances in Machine Learning and Intelligent Cyber Security (DYNAMICS) Workshop}{2019}{Puerto Rico, USA}

\begin{document}
\title{DeepCloak: Adversarial Crafting As a Defensive Measure to Cloak Processes}

\author{Mehmet Sinan \.{I}nci}
\affiliation{Intel Corporation}
\email{mehmet.inci@intel.com}
\author{Thomas Eisenbarth}
\affiliation{University of L\"ubeck}
\email{thomas.eisenbarth@uni-luebeck.de}
\author{Berk Sunar}
\affiliation{Worcester Polytechnic Institute}
\email{sunar@wpi.edu}

\begin{abstract}
Over the past decade, side-channels have proven to be significant and practical threats to modern computing systems. Recent attacks have all exploited the underlying shared hardware. While practical, mounting such a complicated attack is still akin to listening on a private conversation in a crowded train station. The attacker has to either perform significant manual labor or use AI systems to automate the process. The recent academic literature points to the latter option. With the abundance of cheap computing power and the improvements made in AI, it is quite advantageous to automate such tasks. By using AI systems however, malicious parties also inherit their weaknesses, most notably the vulnerability to adversarial samples.

In this work, we propose the use of adversarial learning as a defensive tool to obfuscate and mask side-channel information. We demonstrate the viability of this approach by first training CNNs and other machine learning classifiers on leakage trace of different processes. After training a highly accurate model (99+\% accuracy), we test it against adversarial learning. We show that through minimal perturbations to input traces, the defender can run as an attachment to the original process and cloak it against a malicious classifier.

Finally, we investigate if an attacker can use adversarial defense methods, adversarial re-training and defensive distillation to protect the model. Our results show that even in the presence of an intelligent adversary that employs such techniques, adversarial learning methods still manage to successfully craft perturbations hence the proposed cloaking methodology succeeds.

\end{abstract}	

%\keywords{Machine Learning, Deep Learning, Adversarial Learning, Adversarial Defenses, Side-channel Leakage}
\maketitle

%%%%%%%%%%%%%%%%%%%%%%%%
\section{Introduction}
%%%%%%%%%%%%%%%%%%%%%%%%

Deep learning (DL) has proven to be a very powerful tool for a variety of tasks like handwritten digit recognition~\cite{deep_feats}, image classification and labeling~\cite{deep_diabetic_eye_paper, deep_lung_cancer_paper, deep_breast_cancer_paper}, speech recognition~\cite{deep_speech_paper}, lip reading~\cite{deep_lip_reading_paper}, verbal reasoning~\cite{deep_iq2}, playing competitive video games~\cite{deep_starcraft_paper, deep_smash2} and even writing novels~\cite{deep_romance_novel}. As with any booming technology, it is also adopted by malicious actors e.g.,posting to internet boards to shift public opinion by social engineering. One of the latest known examples of this is the millions of AI-generated comments on the FCC net-neutrality boards~\cite{fcc_ars, fcc_bloomberg}. These AI-crafted posts were semantically sound and not easy to detect as fake. Another example of malicious use of AI is for spam and phishing attacks. It is now possible to craft custom phishing e-mails using AI with higher `yield' than human crafted ones~\cite{AIspam, AIphishing}. %Abundance of cheap computing power makes the bulk creation of such e-mails easier than ever. 
AI system can even participate in hacking competitions where human creativity and intuition was thought to be irreplaceable. In 2016, DARPA sponsored a hacking competition for AI systems where the task was to find and fix vulnerabilities in computer systems within a given time period~\cite{ai_hacking}. According to a survey among cybersecurity experts, the use of AI for cyber attacks will become more common with time~\cite{ai_cyber}.

%Here, we tackle this challenge for a specific domain, side-channel attacks ). 

Classical side-channel attacks (SCA) deal with bulk amounts of noisy data that require human interpretation and intuition to process. Such task are perfectly suited for AI systems and it is reasonable that malicious parties are aware of this opportunity. The academic literature already shows the use of AI for processing side-channel leakage. In 2011 Hospodar et al.~\cite{hospodar2011machine} demonstrated the first use of machine learning, LS-SVM specifically, on a power SCA on AES and showed that the ML approach yields better results than the traditional template attacks. Later, Heuser et al.~\cite{heuser2012intelligent} showed the superiority of multi-class SVM for noisy data in comparison to the template attacks. Martinasek et al.~\cite{martinasek2013innovative, martinasek2013optimization} showed that artificial neural networks can recover AES keys from power measurements with a success rate of 96\%. In 2015 Beltramelli~\cite{beltramelli2015deep} used LSTM to collect meaningful keystroke data via motions of the smart watch user. In 2016, Maghrebi et al.~\cite{maghrebi2016breaking} compared four DL based techniques with template attacks to attack an unprotected AES implementation using power consumption and showed that CNN outperforms template attacks. Finally in 2017, Gulmezoglu et al.~\cite{gulmezoglu2017perfweb} showed that machine learning can be used to extract meaningful information from cache side-channel leakage to recover web traffic of users.

A straightforward countermeasure against SCAs is to drown the sensitive computation leakage in noise to cloak it. However, this defense has proven to be ineffective in addition to being computationally expensive with significant performance overhead. In this study, we argue that we can do much better than random noise and craft much smaller noise by using adversarial learning (AL). By using AL, we achieve a stronger cloaking effect using smaller changes to the trace hence minimal overhead to the system. Also, the proposed defense does not require redesigning the software or the hardware stacks. The proposed framework can be deployed as an opt-in service that users can enable or disable at wish, depending on their privacy needs at the time.

In summary, attacking machine learning is easier than defending it~\cite{attacking_ml} and if used strategically in a non-traditional way i.e., as a defensive countermeasure, AL against malicious parties with AI capabilities can be quite advantageous. In this work, we expand this idea and show that AL is indeed a useful defensive tool to cloak private processes from AI capable adversaries.

%%%%%%%%%%%%%%%%%%%%%%%%%%%%%%%%%%%%%%%%%%
\subsection*{Our Contribution}
%%%%%%%%%%%%%%%%%%%%%%%%%%%%%%%%%%%%%%%%%%
   
In this work, we propose a framework and explore the necessary steps to cloak processes against SCAs as well as the defenses a malicious party can use. More specifically in this paper we;
\begin{itemize}
	\item show how to profile crypto processes with high accuracy via their side-channel leakage using deep learning and various classical machine learning models. We classify 20 types of processes using readily available, high resolution Hardware Performance Counters (HPC). Further, we investigate the effect of parameter choices like the number of features, samples and data collection intervals on the accuracy of such classifiers.
	\item present the use of AL methods to craft perturbations and add them to the system hardware trace to cloak the side-channel leakage of private processes. We show that this is a strong defense against an attacker using DL classifiers.
	\item test and quantify the efficiency of different AL methods and present the accuracy and applicability of each attack.
	\item show that even when adversarial defense methods adversarial re-training or defensive distillation is employed by the attacker, adversarial perturbations still manage to cloak the process.
\end{itemize}

%%%%%%%%%%%%%%%%%%%%%%%%%%%%%%%%%%%%%%%%%%%%
\section{Background}\label{sec:bac}
%%%%%%%%%%%%%%%%%%%%%%%%%%%%%%%%%%%%%%%%%%%%

In this section, we provide the necessary background information to better understand the attack, adversarial sample crafting as a countermeasure and the improved attack. More specifically, we go over micro-architectural attacks, hardware performance counters, convolutional neural networks (CNNs), and AL attacks.

%%%%%%%%%%%%%%%%%%%%%%%%%%%%%%%%%%%%%%%%%%%
\subsection{Micro-architectural Attacks}
%%%%%%%%%%%%%%%%%%%%%%%%%%%%%%%%%%%%%%%%%%%

Over the last decade, there has been a surge of micro-architectural attacks. Low-level hardware bottlenecks and performance optimizations have shown to allow processes running on shared hardware to influence and retrieve information about one another. For instance, cache side-channel attacks like Prime\&Probe and Flush +Reload exploit the cache and memory access time difference to recover fine-grain secret information and even recover secret crypto keys~\cite{percival2005cache, osvik2006cache, yaromFnR, gullasch2011cache, Bernstein05cache-timingattacks, zhang2014cross, waitaminute, inci2016cache, gruss2015cache, lipp2016armageddon, pessl2016drama, gruss2016flush+}. In these works, the attacker exploits micro-architectural leakages stemming from memory access time variations, e.g.,when the data is retrieved from small but faster caches as opposed to slower DRAM memory. 

\subsection{Hardware Performance Counters}
%%%%%%%%%%%%%%%%%%%%%%%%%%%%%%%%%%%%%%%%%
 
Hardware Performance Counters (HPCs) are special purpose registers that provide low-level execution metrics directly from the CPU. This low-level information is particularly useful during software development to detect and mitigate performance bottlenecks before deployment. For instance, low number of cache hits and high cache misses indicate an improperly ordered loop. By re-ordering some operations, a developer can significantly improve the performance. %While there are many different HPCs, availability of a HPC depends on the specific CPU model. Moreover, the number of HPCs that can be monitored simultaneously depends both on the CPU model and the selected HPCs. Since some HPCs are derived from others, their use puts additional limitations to the monitoring process.

HPCs are also useful to obtain system health check and/or anomaly detection in real-time. For instance, in~\cite{alamperformance} Alam et al. leverages \textit{perf\_event} API to detect micro-architectural side-channel attacks. In 2009, Lee et al.~\cite{zhang2016cloudradar} showed that HPCs can be used on cloud systems to provide real-time side-channel attack detection. In~\cite{samira_cacheshield_sca_protection, samira_cacheshield2_sca_detection}, researchers used HPCs to detect cache attacks. Moreover, Allaf et al.~\cite{ML_SCA_detection} used a neural network, decision tree, and kNN to specifically detect Flush+Reload and Prime\&Probe attacks on AES. Moreover, researchers have shown that by using the fine-grain information provided by HPCs, it is possible to violate personal privacy as well. In~\cite{gulmezoglu2017perfweb}, Gulmezoglu et al. showed that HPC traces can be used to reveal the visited websites in a system using variety of ML techniques such as auto-encoder, SVM, kNN and decision trees and works even on privacy conscious Tor browser. More applications of HPCs in cyber security countermeasures can be found in~\cite{foreman2018survey}.

% removed due to space constraints
%\subsection{Preventing Micro-architectural Leakage}
%Side-channel leakages through micro-architectural features, whether measured via cache access profiles, by monitoring branch predictors, or directly via HPCs, can be prevented. The most effective technique is constant-time implementation, where execution behavior must be independent of sensitive inputs. Several tools to validate constant-time properties have been proposed, e.g.,Langley's \texttt{ctgrind}, \texttt{ct-verif}~\cite{almeida2016verifying} and \texttt{CacheD}~\cite{wang2017cached}. Raccoon~\cite{rane2015raccoon} automates the enforcement of a constant-time control flow. Yet, the adoption rate of constant-time implementation is low even for cryptographic libraries, and probably not even considered for other applications. Besides requiring increased development effort, constant-time implementations often have significantly decreased average-case performance. Alternatives that reduce side-channel leakage at lower overheads use sophisticated randomization techniques, e.g.,software diversity~\cite{crane2015thwarting}, where programs randomly choose from various different implementations of the same functionality or simply access random memory locations during run time~\cite{zhang2013duppel}. While often more efficient than constant-time code, none of the designs are optimized to minimize the overhead while maximizing the obfuscation effect of the countermeasure. 

\vspace{-5pt}
%%%%%%%%%%%%%%%%%%%%%%%%%%%%%%%%%%%%%%%%%
\subsection{Convolutional Neural Networks}
%%%%%%%%%%%%%%%%%%%%%%%%%%%%%%%%%%%%%%%%%

Convolutional Neural Networks (CNN) is a supervised feed-forward artificial neural network architecture. %The supervised learning simply means that the data used to train the model is labeled.
%Training a CNN requires minimal human intervention and is easily automatable. 
One important aspect of CNNs is that they don't saturate easily and can reach high accuracy with more training data. Also, unlike the classical ML methods, CNNs do not require data features to be identified and pre-processed before training. Instead, CNNs discover and learn relevant features in the data without human intervention, making them very suitable for automated tasks. %The fact that features aoffers great flexibility and makes CNNs the go-to classifier for processing large amounts of unlabeled data. The disadvantage of the CNN on the other hand is the need for large amounts of training data.% and the computationally expensive training process compared to the classical ML models. 
In the past decade, CNNs surpassed humans in many tasks that were considered nearly impossible to automate. This breakthrough is fueled by the rapid increase in GPU powered parallel processing power and the advancements in deep learning. CNNs have been successfully applied to image, malware and many other classification problems. Training a CNN model is done in 3 phases. First, the labeled dataset is split into three parts; training, validation and test. The training data is fed to the CNN with initial hyper-parameters and the classification accuracy is measured using the validation data. Guided by the validation accuracy results, the hyper-parameters are updated to increase the accuracy of the model while maintaining its generality. After the model achieves the desired validation accuracy, it is tested with the test data and the final accuracy of the model is obtained.

\vspace{-5pt}
%%%%%%%%%%%%%%%%%%%%%%%%%%%%%%%%%%%%%%%%%
\subsection{Adversarial Learning}
%%%%%%%%%%%%%%%%%%%%%%%%%%%%%%%%%%%%%%%%%

AL is a sub-field of machine learning (ML) that studies the robustness of trained models under adversarial settings. The problem stems from the underlying assumption that the training and the test data comes from the same source are consistent in their features. Studies have shown however that by introducing some small external noise or in this context what is commonly referred to as~\texttt{adversarial perturbations}, it is possible to craft adversarial samples and manipulate the output of ML models. In other words, by carefully crafting small perturbations, one can push a test sample from the boundaries of one class to another. Due to the mathematical properties of the high-dimensional space that the classifier operates in, this modification can be very small. AL refers to the group of techniques that are used to perturb test samples to classifiers and force misclassification. While there are many different methods of crafting such perturbations, ideally they are desired to be minimal and not easily detectable.

Adversarial attacks on classical ML classifiers (under both white-box and black-box scenarios) have been known for quite some time~\cite{lowd2005adversarial, laskov2010machine, huang2011adversarial, biggio2011support, zhou2012adversarial, biggio2012poisoning, biggio2013evasion, biggio2014security}. However, it was Szegedy et al.~\cite{szegedy2013intriguing} that first introduced AL attacks on DNNs. The 2013 study showed that very small perturbations that are indistinguishable to human eye can indeed fool CNN image classifiers like ImageNet. The perturbations in the study are calculated using the technique called Limited-memory Broyden-Fletcher-Goldfarb-Shanno (L-BFGS). This algorithm searches in the variable space to find parameter vectors (perturbation) that can successfully fool the classifier. Later in 2014, Goodfellow et al.~\cite{goodfellow2014explaining} improved the attack by using the Fast Gradient Sign Method (FGSM) to efficiently craft minimally different adversarial samples. Unlike the L-BFGS method, the FGSM is computationally conservative and allows much faster perturbation crafting. In 2016, Papernot et al.~\cite{papernot2016limitations} further improved upon Goodfellow's FGSM by using Jacobian Saliency Map Attack (JSMA) to craft adversarial samples. Unlike the previous attacks, JSMA does not modify randomly selected data points or pixels in an image. Instead, it finds the points of high importance with regards to the classifier decision and then modifies these specific pixels. These points are found by taking the Jacobian matrix of the loss function given a specific input sample, allowing an attacker to craft adversarial samples with fewer modifications.

In 2016~\cite{kurakin2016adversarial, liu2016delving, papernot2016transferability}, multiple new adversarial attacks were discovered. Moreover, the research showed that these adversarial samples are transferable i.e., perturbations that can fool a model can also work on other models trained on the same task. In~\cite{papernot2017practical}, Papernot et al. showed that adversarial attacks can also succeed under the black-box attack scenario where an attacker has only access to the classification labels. In this scenario, the attacker has no access to the model parameters such as weights, biases, classification confidence or the loss, therefore, cannot directly compute or use the gradients to craft a perturbation. Instead, the attacker uses the target model as an oracle that labels the inputs and then uses these labeled images to train her own classifier. Authors demonstrated the feasibility of the attack on MetaMind and Deep Neural Network (DNN) classifiers hosted by Amazon and Google. With 84.24\%, 96.19\% and 88.94\% misclassification rates respectively, they were able to fool the targeted classifiers.

In~\cite{dang2017evading}, researchers have shown that by iteratively morphing a structured input, it is possible to craft adversarial samples under black-box attack scenario. Authors have implemented the attack against a PDF malware classifier and have reported 100\% evasion rate. Moreover, the study acknowledges the fact that black-box attack model has a cost of obtaining labeled data from observations and defines and uses a cost function that takes into account the number of observations. The attack works by adding and/or removing compilable objects to the PDF. Black-box scenario does not assume to obtain confidence scores from the model under attack, only the class output. In summary, the AL is an active research area with plethora of new attacks, defenses and application cases emerging daily~\cite{tramer2017ensemble, eykholt2017note, meng2017magnet, carlini2017towards, su2017one}.

In addition to attack classifiers, adversarial learning have also been used to provide privacy for streaming traffic and facial recognition databases~\cite{zhang2019statistical, oh2017adversarial}. In contrast to these works, DeepCloak uses adversarial learning to craft additional traffic on micro-architectural level to mask the overall side-channel leakage.

%%%%%%%%%%%%%%%%%%%%%%%%%%%%%%%%%%%%%%%
\section{Methodology}\label{sec:methodology}
%%%%%%%%%%%%%%%%%%%%%%%%%%%%%%%%%%%%%%%

Our goal is to show that side-channel classifiers can be successfully stopped using the concept of AL. To validate this assumption, we first train DL-based classifiers using real side-channel data, and show their degradation as the result of AL techniques, even if the DL-based classifier is aware of the AL based cloaking defense. In our experiments, we take the following steps:
\begin{enumerate}
	\item Training the process classifier $\mathbf{C}$ using side-channel leakage $\mathbf{\Omega}$.
	\item Crafting adversarial samples $\delta$ to cloak the user processes and force $C$ to misclassify.
	\item Training a new classifier $C'$ with adversarial defense methods; Defensive Distillation and Adversarial Re-training.
	\item Testing previously crafted adversarial samples $\delta$ against the new classifier $C'$. Also crafting and testing new adversarial samples $\delta '$ against the protected classifier $C'$.
\end{enumerate}

\begin{figure}[h!]
	\centering
	\includegraphics[width=0.9\columnwidth]{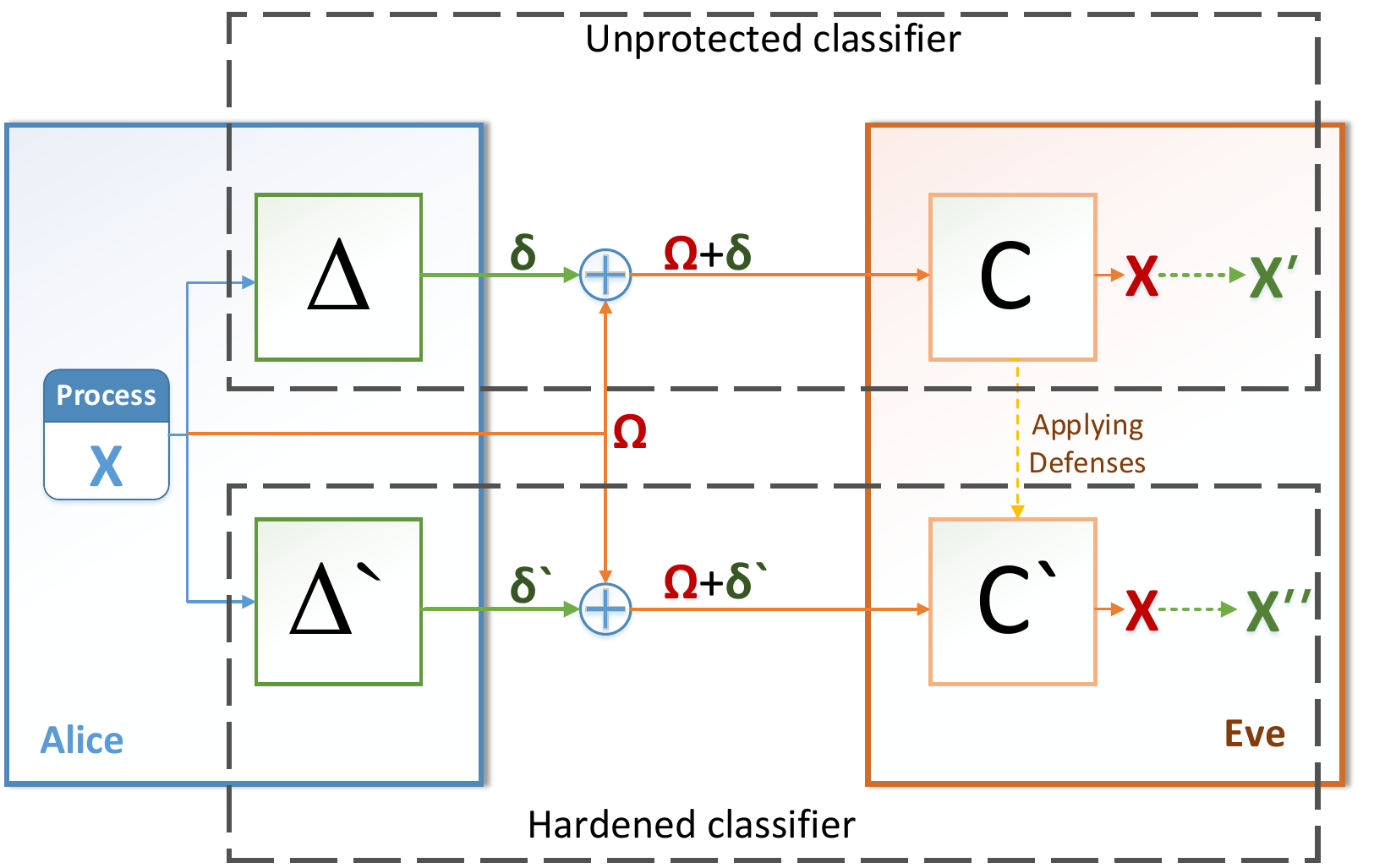}
	\caption{Alice, the defender runs the process X and leaks the $\mathbf{\Omega}$. Eve, the attacker obtains the leakage and identifies the process $\mathbf{X}$ using the classifier, C. Then, Alice crafts the adversarial perturbation $\mathbf{\delta}$ and forces C to misclassify the trace as $\mathbf{X'}$. Eve then trains $\mathbf{C'}$ with adversarial re-training and defensive distillation. Now, Eve can classify $\mathbf{\delta ' + \Omega}$ partially correct. However, when Alice crafts $\mathbf{\delta '}$ against $\mathbf{C'}$, X is again misclassified.}\vspace{-7pt}
	\label{fig:methodology_outline}
\end{figure}

We outline this methodology in Figure~\ref{fig:methodology_outline}. In the first stage, Alice the defender runs a privacy sensitive process $X$. The eavesdropper Eve collects the side-channel leakage $\Omega$ and feeds it into her classifier $C$ and discovers what type of process X is. Then in stage 2, Alice cloaks her process by crafting the adversarial sample $\delta$. When faced with this adversarial sample, Eve's classifier $C$ fails and misclassifies the leakage trace as $X'$. In the third stage, Eve trains a new classifier $C'$ using \emph{defensive distillation} and \emph{adversarial re-training} to protect it from misclassification cause by the adversarial perturbation $\delta$. In the final stage, Alice first tests previously crafted adversarial samples against Eve's protected classifier $C'$. Then, Alice updates her adversarial sample crafting target to fool $C'$ rather than the original classifier $C$.

We apply this methodology to a scenario where a malicious party trains a CNN to classify running processes using the HPC trace as the input. This information is extremely useful to the attacker since it helps to choose a specific attack or pick a vulnerable target among others. Once a a target is found, an attacker can perform micro-architectural or application specific attacks. To circumvent this information leakage and protect processes, the defender attempts to mask the process signature. Ideally, the masking is minimal and does not interfere with the running process.

Specifically, in our methodology, the attacker periodically collects 5 HPC values over 10 msec total with 10 usec intervals, resulting in total of 5000 data points per trace. Later, trace is fed into classical ML and DL classifiers. In this section, we explain our choice of the specific HPCs, the application classifier design and implementation details, the AL attacks applied to these classifiers and finally test the efficiency of adversarial defenses against our cloaking method.

%%%%%%%%%%%%%%%%%%%%%%%%%%%%%%%%%%%%%%%%%%%%%%%%%%%%%%%%%%
\subsection{HPC Profiling}
%%%%%%%%%%%%%%%%%%%%%%%%%%%%%%%%%%%%%%%%%%%%%%%%%%%%%%%%%%

HPCs are special purpose registers that provide detailed information on low-level hardware events in computer systems. These counters periodically count specified event like cache accesses, branches, TLB misses and many others. This information is intended to be used by developers and system administrators to monitor and fine-tune performance of applications. The availability of a specific counter depends on the architecture and model of the CPU. Among many available HPCs, we have selected the following 5 for the classification task;

\begin{enumerate}[noitemsep]
    \item \textbf{Total Instructions:} the total number of retired i.e., executed and completed CPU instructions.
    \item \textbf{Branch Instructions:} the number of branch instructions (both taken and not taken).
    \item \textbf{Total Cache References:} the total number of L1, L2, and L3 cache hits and misses.
    \item \textbf{L1 Instruction Cache Miss:} the occurrence of L1 cache instruction cache misses.
    \item \textbf{L1 Data Cache Miss:} the occurrence of L1 cache data cache misses.
\end{enumerate}

We have selected these HPCs to cover a wide variety of hardware events with both coarse and fine-grain information. For instance, the~\texttt{Total Instructions} does not directly provide any information about the type of the instructions being executed. However, different instructions execute in varying number of cycles even if the data is loaded from the same cache level. This execution time difference translates indirectly into the total instructions executed in the given time period and hints about the instruction being executed. 

\texttt{Branch Instructions} HPC provides valuable information about the execution flow as well. Whether the branches are taken or not taken, the total number of branches in the executed program remains constant for a given execution path. This constant in the leakage trace helps eliminate noise elements and increases classification accuracy. The~\texttt{Total Cache References} HPC provides similar information to the Branch Instructions HPC in the sense that it does not leak information about the finer details like the specific cache set or even the cache level. However it carries information regarding the total memory access trace of the program. Regardless of the data being loaded from the CPU cache or the memory, the total number of cache references will remain the same for a given process. The~\texttt{L1 Instruction Cache Miss} and the~\texttt{L1 Data Cache Miss} HPCs provide fine-grain information about the~\textit{Cold Start} misses on the L1 cache. Since the L1 cache is small, the data in this cache level is constantly replaced with new data, incrementing these counters. Moreover, separate counters for the instruction and the data misses allows the profiler to distinguish between arithmetic and memory intensive operations and increase accuracy. Finally, all five of the HPCs are interval counters meaning that they count specific hardware events within selected time periods.

%%%%%%%%%%%%%%%%%%%%%%%%%%%%%%%%%%%%%%%%%%%%%%%%%%%%%%%%%%
\subsection{Classifier Design and Implementation}\label{ssec:classDesign}
%%%%%%%%%%%%%%%%%%%%%%%%%%%%%%%%%%%%%%%%%%%%%%%%%%%%%%%%%%

In the first part of the study, we design and implement classifiers that can identify processes using the HPC leakage. To show the viability of such classifier, we chose 20 different ciphers from the OpenSSL 1.1.0 library as the classification target. Note that these classes include ciphers with both very similar and extremely different performance traces e.g.,AES-128, ECDSAB571, ECDSAP521, RC2 and RC2-CBC. Moreover, we also trained models to detect the version of the OpenSSL library for a given cipher. For this task, we used OpenSSL versions 0.9.8, 1.0.0, 1.0.1, 1.0.2 and 1.1.0. %The full list of classified processes is given in Appendix~\ref{app:test_classes}.

\subsubsection{\textbf{Classical ML Classifiers:}} In this study, we refer to non-neural network classification methods as classical ML classifiers. In order to compare and contrast classical ML methods with CNNs, we trained a number of different classifiers using the Matlab Classification Learning Toolbox. The trained classifiers include SVMs, decision trees, kNNs and variety of ensemble methods.

\subsubsection{\textbf{Deep Learning Classifier:}} We designed and implemented the CNN classifier using Keras with Tensorflow-GPU back-end. The model has the total of 12 layers including the normalization and the dropout layers. In the input layer, the first convolution layer, there are a total of 5000 neurons to accommodate the 10 msec of leakage data with 5000 HPC data points. Since the network is moderately deep but extremely wide, we used 2 convolution and 2 MaxPool layers to reduce the number dimensions and extract meaningful feature representations from the raw trace.

In addition to convolution and MaxPool layers, we used batch normalization layers to normalize the data from different HPC traces. This is a crucial step since the hardware leakage trace is heavily dependent on the system load and scales with overall performance. Due to this dependency, the average execution time of a process or parts of a process can vary from one execution to another. Moreover, in the system-wide leakage collection scenario, the model would train over this system load when it should be treated as noise. If not handled properly, the noise and shifts in the time domain results in over-fitting the training data with the dominant average execution time, decreasing the classification rate. By using the batch normalization layer, the model learns the features within short time intervals and the relation between different HPC traces. Finally, the output layer has 20 neurons with softmax activation, each representing a classes of process. To train the model, we use~\emph{Categorical Cross-entropy} loss function with the~\emph{Adam Optimizer}.

%Our CNN classifier is constructed using the layers given below;
%\begin{enumerate}[noitemsep]
%    \item Convolution layer (50, (10,1))
%    \item MaxPool Layer (10,1)
%    \item Batch Normalization Layer
%    \item Dropout Layer (0.25)
%    \item Convolution Layer (100, (10,1))
%    \item MaxPool Layer (10,1)
%    \item Batch Normalization Layer
%    \item Dropout Layer (0.25)
%    \item Flatten Layer
%    \item Dense Layer (400)
%    \item Dropout Layer (0.25)
%    \item Dense Layer (20)
%\end{enumerate}

%%%%%%%%%%%%%%%%%%%%%%%%%%%%%%%%%%%%%%%%%%%%%%%%%%%%%%%%%%
\subsection{Adversarial Learning Attacks}\label{attack:adv_pert}
%%%%%%%%%%%%%%%%%%%%%%%%%%%%%%%%%%%%%%%%%%%%%%%%%%%%%%%%%%

AL remains an important open research problem in AI. Traditionally, AL is used to fool AI classifiers and test model robustness against malicious inputs. In this study however, we propose to use AL as a defensive tool to mask the side-channel trace of applications and protect against micro-architectural attacks and privacy violations. In the following, we explain the specific adversarial attacks that we have used. %Figure~\ref{fig:adv_attacks} shows all 10 adversarial attacks successfully applied to a cat image to fool the ResNet50 image classifier. Each adversarial sample given in the figure successfully fools the classifier. The original, unmodified image is only marginally different than the perturbed ones and therefore is provided in the Appendix~\ref{app:gundi}. 
We consider the following 10 attacks:

\begin{itemize}[leftmargin=*]
 	\item \textbf{Additive Gaussian Noise Attack (AGNA):} Adds Gaussian Noise to the input trace to cause misclassification. The standard deviation of the noise is increased until the misclassification criteria is met. This AL attack method is ideal to be used in the cloaking defense due to the ease of implementation of the all-additive perturbations. A sister-process can actuate such additional changes in the side-channel trace by simply performing operations that increment specific counters like cache accesses or branch instructions.
	\item \textbf{Additive Uniform Noise Attack (AUNA):} Adds uniform noise to the input trace. The standard deviation of the noise is increased until the misclassification criteria is met. Like AGNA, AUNA is easy to implement as a sister-process due to its additive property.
	%\item \textbf{Boundary Attack (BA):} Is introduced by Brendel et al.~\cite{brendel2017decision}. The attack does not require the gradient or the classification confidence value from the model. It works by starting from a large adversarial perturbation and reducing this perturbation within the misclassification boundary. In order to do this, the attacker starts from a perturbed input and performs a random walk on the correct class boundary. Then the adversary takes smaller and smaller steps and progressively reduces the distance to the class boundary hence the perturbation.
	\item \textbf{Blended Uniform Noise Attack (BUNA):} Blends the input trace with Uniform Noise until the misclassification criteria is met.
	\item \textbf{Contrast Reduction Attack (CRA):} Calculates perturbations by reducing the `contrast' of the input trace until a misclassification occurs. In case of the side-channel leakage trace, the attack smooths parts of the original trace and reduces the distance between the minimum and the maximum data points.
	\item \textbf{Gradient Attack (GA):} Creates a perturbation with the loss gradient with regards to the input trace. The magnitude of the added gradient is increased until the misclassification criteria is met. The attack only works when the model has a gradient.
    \item \textbf{Gaussian Blur Attack (GBA):} Adds Gaussian Blur to the input trace until a misclassification occurs. Gaussian blur smooths the input trace and reduces the amplitude of outliers. Moreover, this method reduces the resolution of the trace and cloaks fine-grain leakage.
    \item \textbf{GSA (Gradient Sign Attack)~\cite{goodfellow2014explaining}:} Also called the Fast Gradient Sign Method, the attack has been proposed by Goodfellow et al. in 2014. GSA works by adding the sign of the elements of the gradient of the cost function with regards to the input trace. The gradient sign is then multiplied with a small constant that is increased until a misclassification occurs.
    \item \textbf{L-BFGS-B Attack (LBFGSA)~\cite{LBFGSA_paper}:} The attack utilizes the modified Broyden-Fletcher-Goldfarb-Shanno algorithm, an iterative method for solving unconstrained nonlinear optimization problems, to craft perturbations that have minimal distance to the original trace. The attack morphs the input to a specific class. However, in our experiments, we did not target a specific class and chose random classes as the target.
    \item \textbf{Saliency Map Attack (SMA)~\cite{papernot_saliency}:} Works by calculating the forward derivative of the model to build an adversarial saliency map to detect which input features e.g.,pixels in an image, have a stronger effect on the targeted misclassification. Using this information, an adversary can modify only the features with high impact on the output and produce smaller perturbations.    
	\item \textbf{Salt and Pepper Noise Attack (SPNA):} Works by adding Salt and Pepper noise (also called impulse noise) to the input trace until a misclassification occurs. For images, salt and pepper values correspond to white and black pixels respectively. For the side-channel leakage trace however, these values correspond to the upper and the lower bounds in the trace.
\end{itemize}

%\begin{figure*}[h!]
%	\centering
%	\includegraphics[width=2.0\columnwidth]{images/adv_attacks2.pdf}
%	\caption{Various adversarial attacks applied to an image of a cat against the ResNet50 image classifier. All of the shown perturbations successfully cause misclassification.}
%	\label{fig:adv_attacks}
%\end{figure*}

%%%%%%%%%%%%%%%%%%%%%%%%%%%%%%%%%%%%%%%%%%%%%%%%%%%%%%%%%%
\subsection{Adversarial Learning Defenses}\label{attack:adv_def}
%%%%%%%%%%%%%%%%%%%%%%%%%%%%%%%%%%%%%%%%%%%%%%%%%%%%%%%%%%

In order to see the viability of any defense method that can be used by an attacker against our adversarial perturbations, we have explored two methods: \emph{adversarial re-training} and \emph{defensive distillation} (DD). These defenses are an integral part of this study since an attacker capable of overcoming adversarial perturbation would deem any cloaking mechanism moot.

\subsubsection{\textbf{Gradient Masking:}} The term gradient masking defense has been introduced in~\cite{papernot2017practical} to represent group of defense methods against adversarial samples. The defense works by hiding the gradient information from the attacker to prevent it from crafting adversarial samples. Papernot et al.~\cite{papernot2017practical} however showed that the method fail under the oracle access scenario. An attacker can query the classifier with enough samples to create a cloned classifier. Since the clone and the original classifiers have correlated gradients, the attacker can use the gradient from the clone and craft adversarial samples, bypassing the defense. Due to the known weaknesses and limitations of this defense method, we do not further investigate it in this study.

\subsubsection{\textbf{Adversarial Re-training:}} This defense idea was first proposed by Szegedy et al. in 2013~\cite{szegedy2013intriguing}. Later in 2014, Goodfellow et al.~\cite{goodfellow2014explaining} improved the practicality of the method by showing how to craft adversarial samples efficiently using the Fast Gradient Sign Method. In this defense, the model is re-trained using adversarial samples. By doing so, the model is `vaccinated' against adversarial perturbations and can correctly classify them. In other words, the method aims to teach adversarial perturbations to the model so that it can generalize better and not be fooled by small perturbations. While this method works successfully against a specific type of attack, it has been shown to fail against attack methods that the model was not trained for. Nevertheless, we apply this defense method to our classifiers and investigate its applicability to side-channel leakage classifiers.

\subsubsection{\textbf{Defensive Distillation:}} The DD has been proposed by Papernot et al.~\cite{papernot2016transferability} in 2016 to protect DL models against AL attacks. The goal of this technique is to increase the entropy of the prediction vector to protect the model from being easily fooled. The method works by pre-training a model with a custom output layer. Normally, the softmax temperature is set to be as small as possible to train a tightly fitted, highly accurate model. In the custom layer however, the temperature value is set to a higher value to distill the probability outputs. The first model is trained with the training data using hard labels i.e., the correct class label is set to `1' and all other class labels are set to `0'. After the model is trained, the training samples are fed into it and the probability outputs are recorded as~\emph{soft labels}. Then these soft labels are used to train the second, distilled model with the same training data. This process smooths the model surface on directions that an adversary would use to craft perturbations. This smoothing process increases the perturbation size required to craft an adversarial samples and invalidates some of the previously crafted adversarial samples. This smoothing can be set to different levels by adjusting the temperature value. Note that however, the DD can reduce the classification accuracy significantly if the temperature value is set too high.

%%%%%%%%%%%%%%%%%%%%%%%%%%%%%%
\section{Experiment Setup and Results}\label{sec:results}
%%%%%%%%%%%%%%%%%%%%%%%%%%%%%%

In this section, we give details of our experiment setup, and the results of different adversarial attacks on the crypto-process classifier, and finally present the results of hardened adversarially re-trained and distilled models.

\subsubsection*{\textbf{Experiment Setup:}}\label{sec:exp} DL models perform large amounts of matrix multiplications that can run efficiently in modern GPU systems. For that reason, we used a workstation with two Nvidia 1080Ti (Pascal architecture) GPUs, 20-core Intel i7-7900X CPU and 64 GB of RAM. On the software side, the classifier model is coded using \textit{Keras v2.1.3} with \textit{Tensorflow-GPU v1.4.1} back-end and other Python3 packages such as \textit{Numpy v1.14.0, Pandas, Sci-kit, H5py etc}.

The HPC data is collected from a server with Intel Xeon E5-2670 v2 CPU running Ubuntu 16 LTS. The CPU has 85 HPCs of which 50 are accessible from user-space. To access the HPCs, we had the choice of using~\textit{Perf} and~\textit{PAPI} libraries. Due to the lower sampling rate of Perf, we chose to use the PAPI with~\textit{QuickHPC}~\cite{quickhpc} front-end. QuickHPC is a tool developed by Marco Chiappetta to collect high-resolution HPC data using the PAPI back-end. It is over 30000 times faster than~\textit{perf-stat} and provides an easy to use interface.

%%%%%%%%%%%%%%%%%%%%%%%%%%%%%%%%%%%%%%%%%%%%%%%%%%%%%%%%%%
\subsection{Classification Results}
%%%%%%%%%%%%%%%%%%%%%%%%%%%%%%%%%%%%%%%%%%%%%%%%%%%%%%%%%%
The classifiers are trained to identify 20 classes representing a diverse set of different ciphers of five different versions of OpenSSL, as detailed in Section~\ref{ssec:classDesign}. For training, we split our dataset into three parts as training (\%60), validation (\%20) and test (\%20).
%Unless otherwise stated, the classifier is trained with data from the cipher implementations of OpenSSL 1.1.0, the same version we use to craft adversarial samples against in later experiments.
%Moreover, to show the versatility of the classifier i.e., that it is not limited to a certain process type or a library, we trained models to classify 20 different ciphers from OpenSSL versions 0.9.8, 1.0.0., 1.0.1, 1.0.2 and 1.1.0. Further, we trained models to distinguish between different versions of the same process e.g. OpenSSL 1.0.0 implementation of AES-128-CBC vs any other version of the OpenSSL.

\subsubsection{\textbf{CNN Classifier:}\label{subsec:CNN_res}} 
\begin{figure}[!t]
	\centering
	\includegraphics[width=0.9\columnwidth]{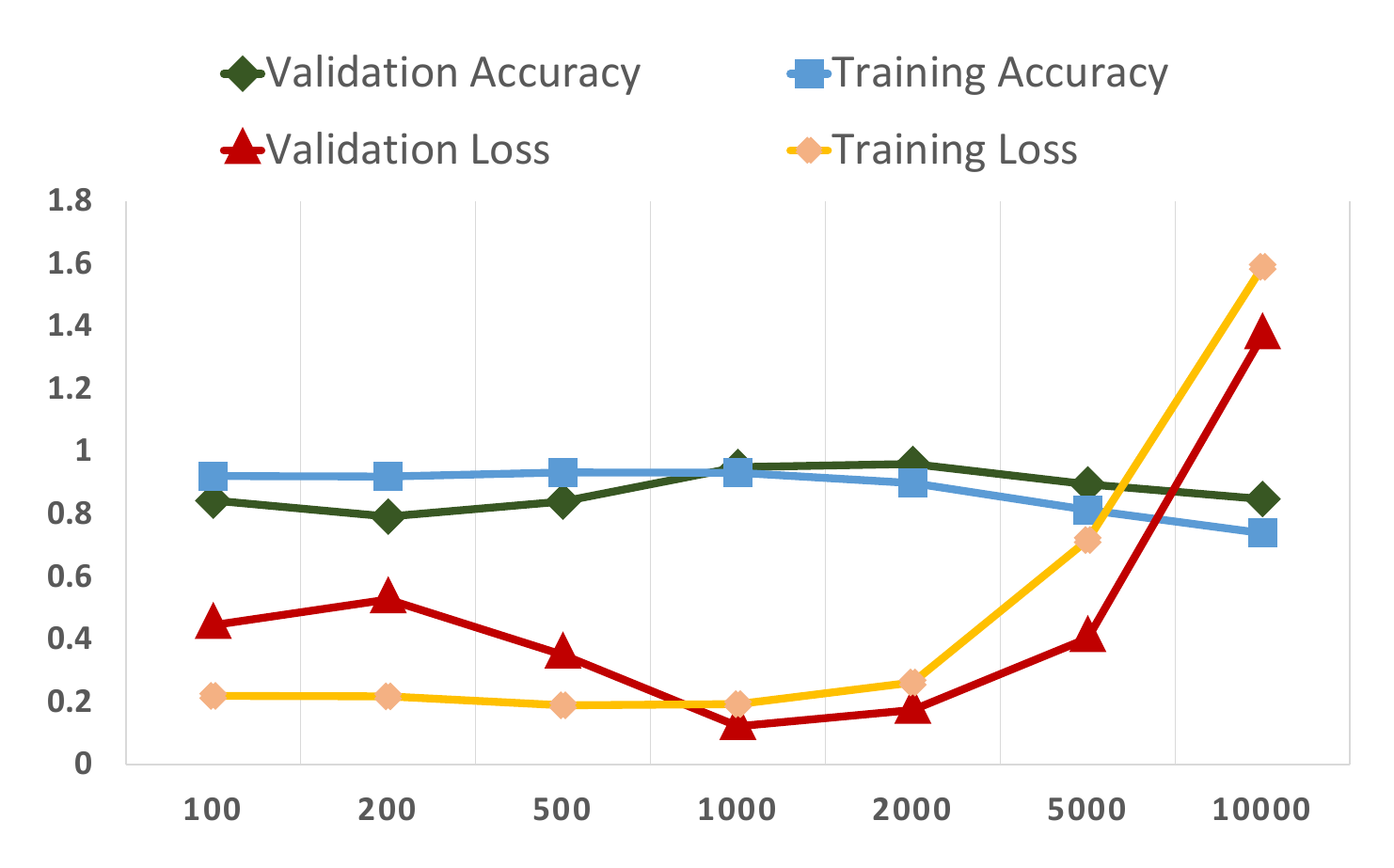}
	\caption{Results for the CNN classifier trained using varying number of features. Models reach highest validation accuracy with 1000 and 2000 features.\vspace{-9pt}}
	\label{fig:CNN_results_feature}
\end{figure}

\begin{figure*}[t!]
	\centering
	\includegraphics[width=0.85\columnwidth]{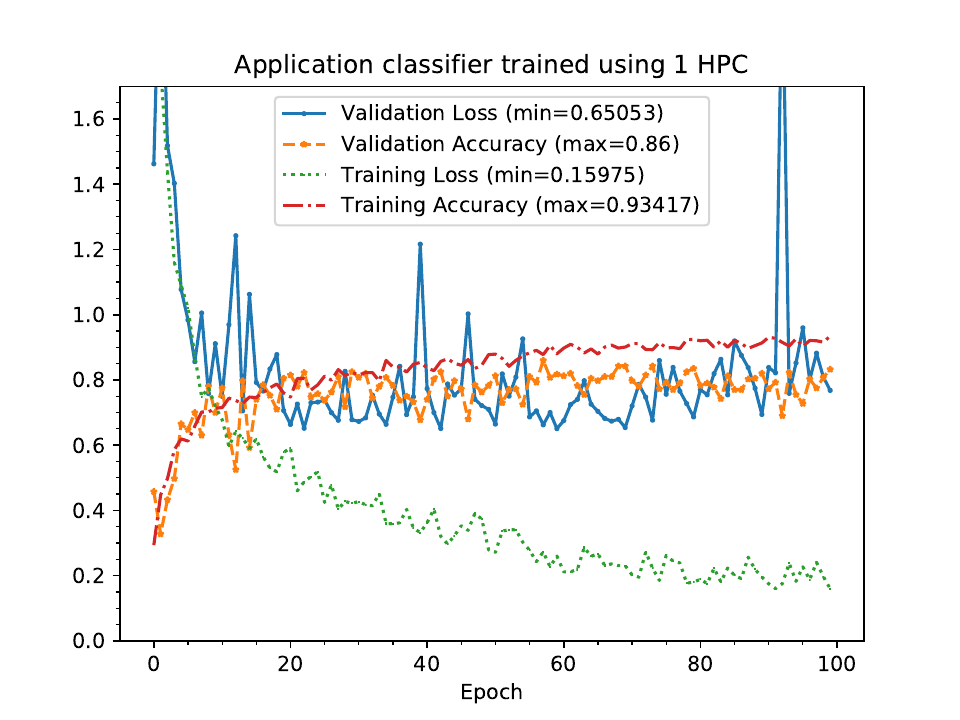}
	\includegraphics[width=0.85\columnwidth]{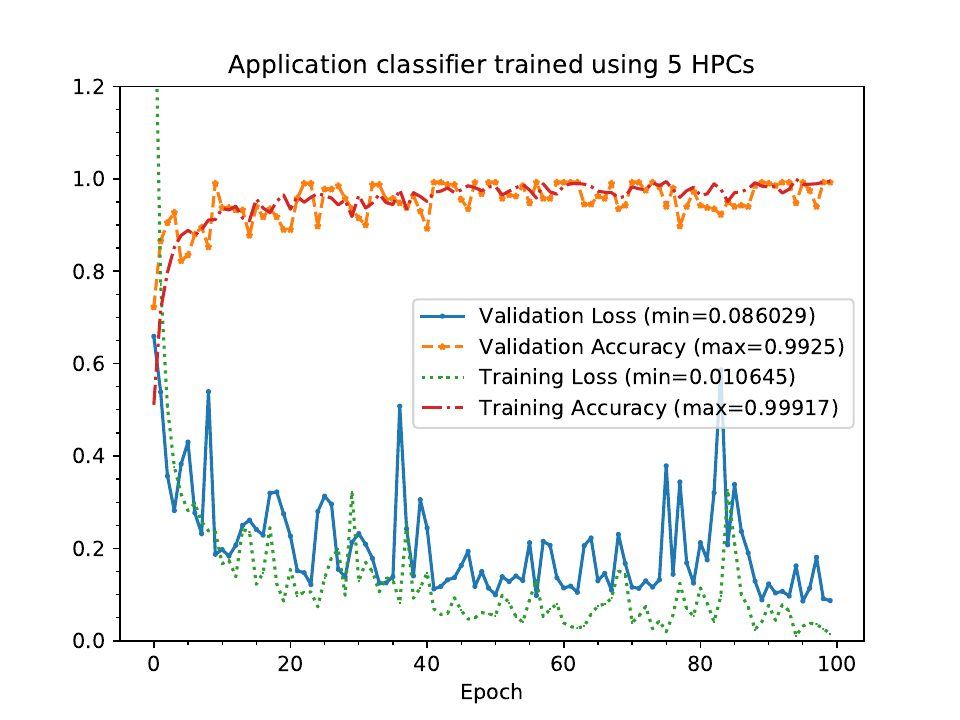}
	\caption{Classification accuracy of models trained using 1 and 5 HPCs. Even using data from a single HPC trace is enough to obtain high accuracy top-1 classification rates, albeit taking longer to train.}
	\label{fig:CNN_results_HPC}
\end{figure*}

For the CNN classifier, we firstly investigated the effect of the number of HPCs collected and trained our models for 100 epochs with data from a varying number of HPCs. Not surprisingly, even using a single HPC, our CNN classifier achieved 81\% validation accuracy by training more epochs. Moreover, we noticed that after the 30th epoch, the model overfitted the training data i.e., the validation accuracy started to drop while the training accuracy kept increasing. When we increased the number of HPCs collected, our models became much more accurate and achieved over 99\% validation accuracy as seen in Figure~\ref{fig:CNN_results_HPC}. Moreover, when we use the data from all 5 HPCs, our model achieved 99.8\% validation accuracy in less than 20 epochs. While our validation accuracy saturates even with only 2 HPCs,~\textit{Total Instructions} and~\textit{Branch Instructions} we have decided to use all 5 of them. We made this decision because in a real-world attack scenario, an attacker might be using any one or more of the HPCs. Since it would not be known which specific hardware event(s) an attacker would monitor, we decided to use all 5, monitoring different low level hardware events to provide a comprehensive cloaking coverage. To find the optimum number of features per HPC, we have trained multiple models with using various number of features. As shown in Figure~\ref{fig:CNN_results_feature}, the validation accuracy saturates at 1000 and 2000 features, validation loss drops after 1000 features. For this reason, we chose to use 1000 features for our experiments. 

%\vspace{-5pt}

\begin{figure*}[t!]
	\centering
	\includegraphics[width=0.85\columnwidth]{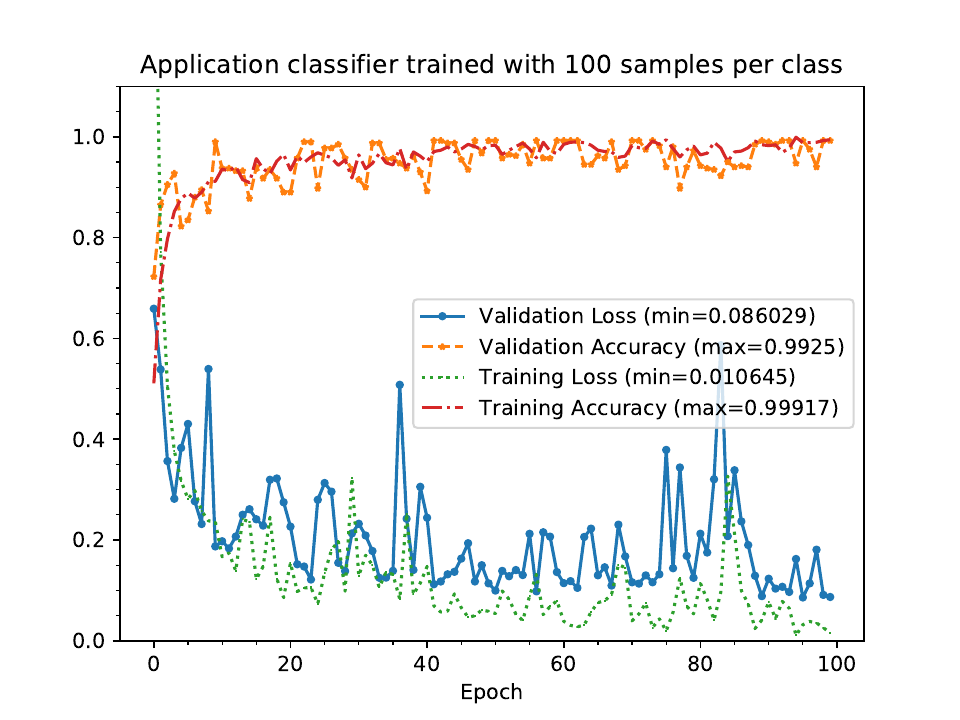}
	\includegraphics[width=0.85\columnwidth]{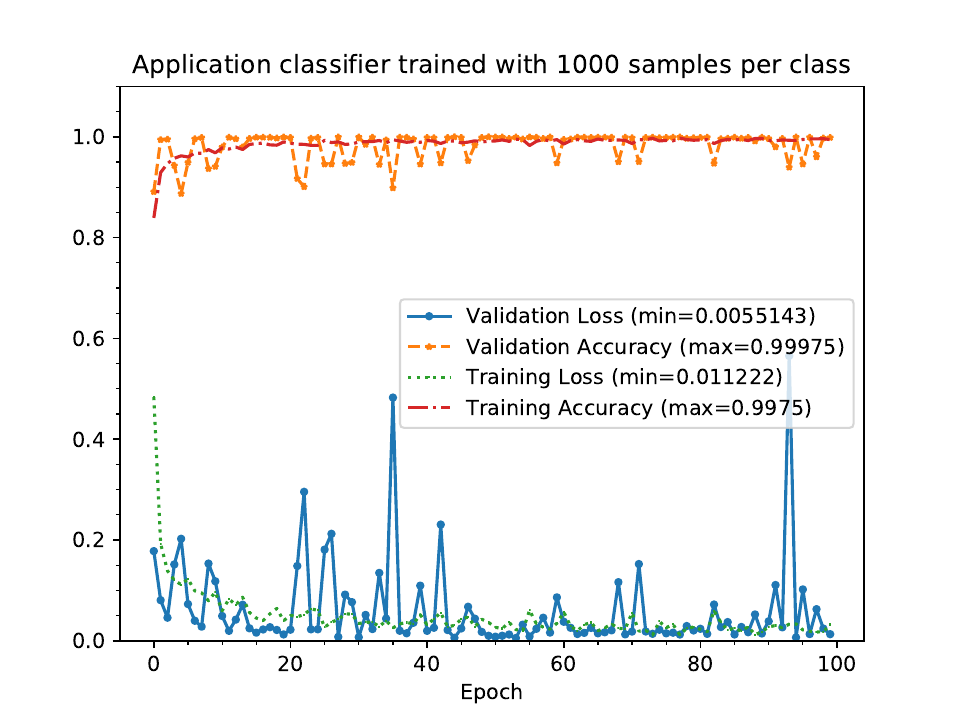}
	\caption{Classification accuracy of models trained using 100 and 1000 samples per class. Both models reach 99\% accuracy in 20 and 40 epochs of training respectively.}
	\label{fig:CNN_results_sample}
\end{figure*}

%After deciding to use data from 5 HPCs, 
Further, we investigated how the number of training samples affect the validation accuracy. For that, we have trained 6 models with a varying number of training samples. For the first model, we have used only 100 samples per class (2000 samples in total) and later on trained models with 300, 1000, 3000, 10000 and 30000 samples per class. In the first model, we achieved 99.8\% validation accuracy after 40 epochs of training. When we trained models with more data, we have reached similar accuracy levels in much fewer epochs. To make a good trade-off between the dataset size and training time, we have opted to use 1000 samples per class. This model reached 100\% accuracy with 20 epochs of training as shown in Figure~\ref{fig:CNN_results_sample}. Finally, our last model achieved 100\% accuracy just after 4 epochs when trained with 30000 samples per class. %Additional results on varying number samples are presented in Appendix~\ref{appfig:CNN_results_sample}.

\begin{figure*}[t!]
	\centering
	\includegraphics[width=0.85\columnwidth]{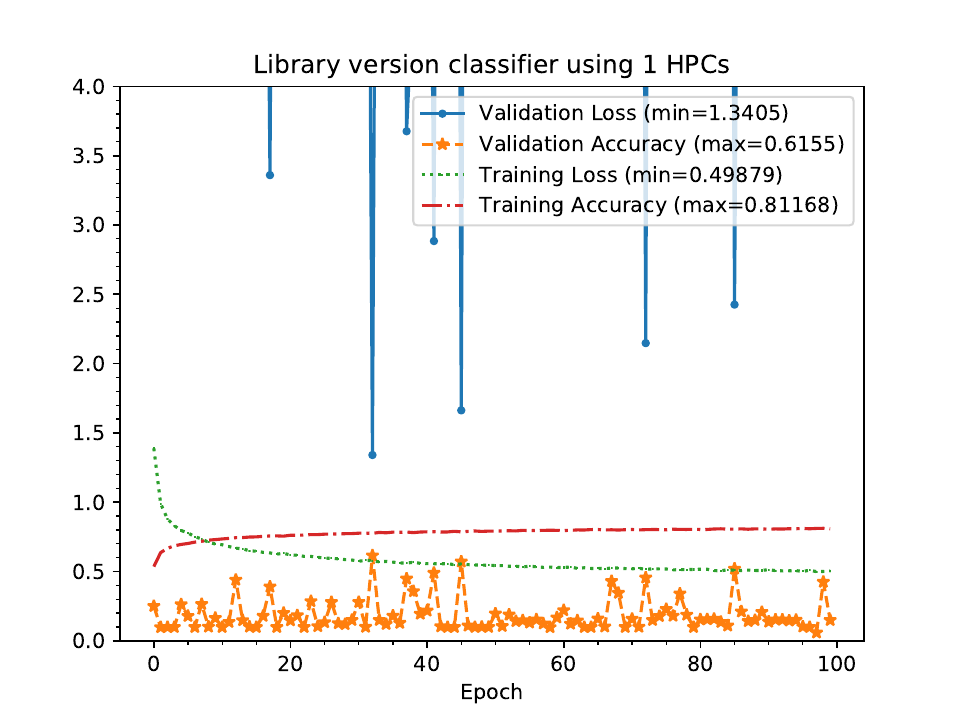}
	\includegraphics[width=0.85\columnwidth]{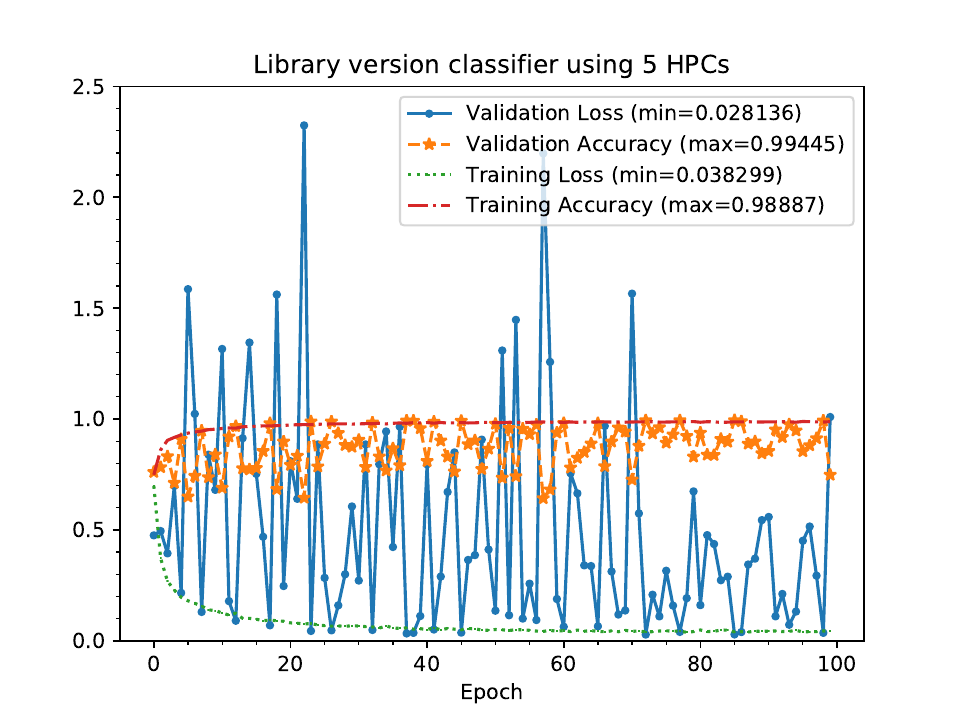}
	\caption{Library version classifier accuracy for models trained using 1 and 5 HPCs. Accuracy of the former model saturate at 61\% while the latter reach 99\%.}
\label{fig:CNN_results_lib}
\end{figure*}

We also show that in addition to detecting the process type, we can also distinguish between different versions of OpenSSL. For each of the 20 analyzed ciphers, we built classifiers to identify the library version. Figure~\ref{fig:CNN_results_lib} presents the classification results of two models trained using 1 and 5 HPC traces respectively. %More results with 2,3 and 4 HPCs are presented in Appendix~\ref{app:class_results}. 
As cipher updates between versions can be very small, the added information from sampling several HPCs is essential for high classification rates, as shown in the results.

\begin{table}[t]
	\centering
	\caption{Application classification results for the classical ML classifiers with and without PCA feature reduction.}
	\label{table:ML_results}
	\resizebox{0.8\columnwidth}{!}{
		\begin{tabular}{@{}lcc@{}}
			\toprule
			\textbf{Classification Method}     & \textbf{Without}  & \textbf{With PCA} \\      		   & \textbf{PCA}		   & \textbf{(99.5\% variance)} \\ \midrule
			Fine Tree              & 98.7      & 99.9                \\
			Medium Tree            & 85.4      & 94.8                \\
			Coarse Tree            & 24.9      & 25                  \\
			Linear Discriminant    & 99.6      & 99.7                \\
			Quadratic Discriminant & N/A    	   & 99.4                \\
			Linear SVM             & 99.9      & 98.2                \\
			Quadratic SVM          & 99.9      & 96.9                \\
			Cubic SVM              & 99.9      & 94.3                \\
			Fine Gaussian SVM      & 40        & 88.2                \\
			Medium Gaussian SVM    & 98.3      & 92.1                \\
			Coarse Gaussian SVM    & 99.7      & 13.4                \\
			Fine kNN               & 96.8      & 11.1                \\
			Medium kNN             & 94.9      & 7.8                 \\
			Coarse kNN             & 85.5      & 5.2                 \\
			Cosine kNN             & 92.5      & 19.6                \\
			Cubic kNN              & 85.2      & 7.7                 \\
			Weighted kNN           & 95.9      & 8.3                 \\
			Boosted Trees          & 99.2      & 99.8                \\
			Bagged Trees           & 99.9      & 94.8                \\
			Subspace Discriminant  & 99.8      & 99.7                \\
			Subspace kNN           & 84.8      & 88.1                \\
			RUSBoosted Trees       & 76        & 92.8             \\\midrule
			\textbf{Best}          & \textbf{99.9}      & \textbf{99.9}                \\ \bottomrule
		\end{tabular}	
	}
\end{table}

\subsubsection{\textbf{Classical ML Methods:}} In our training of ML classifiers, the first challenge was the fact that the side-channel leakage data is extremely wide. We have chosen to train our models using 1000 data points per HPC with 5 HPCs monitored simultaneously. This parameter selection is done empirically to provide wide cloaking coverage and train highly accurate models as explained in Section~\ref{subsec:CNN_res}. Using 1000 data points with 10 usec intervals per HPC allowed us to obtain high quality data in a short observation window. Nevertheless, 5000 dimensions is unusually high for classifiers, especially considering that we are training multi-class classifiers with 20 possible classes.

In order to find optimal settings for the hardware leakage trace, we tried different parameters with each classifier. For instance in the case of decision trees, we have trained the `Fine Tree', `Medium Tree' and `Coarse Tree' classifiers. The difference between these classifiers is that respectively they allow 5, 20, and 100 splits (leaves in the decision tree) to better distinguish between classes. For the case of Gaussian SVM, fine, medium and coarse refers to the kernel scale set to sqrt(P)/4, sqrt(P) and sqrt(P)*4 respectively. As for the kNN, the parameters refer to the number of neighbors and the different distance metrics. %More detailed description of these classifiers is provided in Appendix~\ref{app:classical_ml}.

Results for the classical ML classifiers are given in Table~\ref{table:ML_results}. Classic ML algorithms achieve very high success rates for the given classification task. The trained models can in fact classify running processes by using their HPC traces. Note that the Quadratic Discriminant did not converge to a solution without the PCA application hence no score is given in the table.

%%%%%%%%%%%%%%%%%%%%%%%%%%%%%%%%%%%%%%%%%%%%%%%%%%%%%%%%%%
\subsection{AL on the Unprotected Model} 
%%%%%%%%%%%%%%%%%%%%%%%%%%%%%%%%%%%%%%%%%%%%%%%%%%%%%%%%%%
Next, we crafted adversarial perturbations for the unprotected classifiers by using and adapting the publicly available Foolbox~\cite{rauber2017foolbox} library to our scenario. The library provides numerous adversarial attacks and provides an easy to use API. For a selected attack, Foolbox crafts necessary perturbations on a given sample and classifier model pair to `fool' the given model. Detailed information about these attacks can be found in Section~\ref{attack:adv_pert}.

\begin{table*}[th]	
	\centering
	\caption{AL results of the unprotected and the hardened (Adversarial Re-training) classifiers. Adversarial samples can fool both models and new perturbations can be successfully crafted against the hardened model. Albeit lowering the misclassification confidence up to 29\%, re-training does not protect against AL.}
	\label{table:retrained_adversarial}
	\resizebox{0.8\textwidth}{!}{%
		\begin{tabular}{@{}lcccccc@{}}
			\toprule
			& \multicolumn{3}{c|}{\textbf{Unprotected Classifier}} & \multicolumn{3}{c}{\textbf{Hardened Classifier (Adv. Re-training)}} \\ \midrule
			\textbf{\begin{tabular}[c]{@{}l@{}}Adversarial\\ Attack\end{tabular}} & \textbf{\begin{tabular}[c]{@{}c@{}}Untouched Sample\\ Classification\\ Confidence\end{tabular}} & \textbf{\begin{tabular}[c]{@{}c@{}}Perturbed Sample\\ MisClassification\\ Confidence\end{tabular}} & \textbf{\begin{tabular}[c]{@{}c@{}}Perturbation\\ Size (MAD)\end{tabular}} & \textbf{\begin{tabular}[c]{@{}c@{}}Untouched Sample\\ Classification\\ Confidence\end{tabular}} & \textbf{\begin{tabular}[c]{@{}c@{}}Perturbed Sample\\ MisClassification\\ Confidence\end{tabular}} & \textbf{\begin{tabular}[c]{@{}c@{}}Perturbation\\ Size (MAD)\end{tabular}} \\ \midrule
			\textbf{AGNA} & 99 & 96 & 0.00294 & 92 & 82 & 0.00294 \\
			\textbf{AUNA} & 99 & 97 & 0.00292 & 91 & 82 & 0.00332 \\
			%			\textbf{BA} & 99 & 49 & 0.00022 & 93 & 49 & 0.00228 \\
			\textbf{BUNA} & 99 & 99 & 0.05000 & 96 & 93 & 0.05000 \\
			\textbf{CRA} & 99 & 99 & 0.05254 & 97 & 98 & 0.04999 \\
			\textbf{GA} & 99 & 99 & 0.00250 & 88 & 97 & 0.00398 \\
			\textbf{GBA} & 99 & 97 & 0.00080 & 93 & 74 & 0.00071 \\
			\textbf{GSA} & 99 & 99 & 0.00499 & 89 & 97 & 0.00596 \\
			\textbf{LBFGSA} & 99 & 86 & 0.00025 & 89 & 72 & 0.00031 \\
			%			\textbf{RA} & 99 & 94 & 0.00005 & 95 & 48 & 0.00316 \\
			%			\textbf{SLSQPA} & 99 & 100 & 49.66867 & 92 & 100 & 49.66902 \\
			\textbf{SMA} & 99 & 92 & 0.00001 & 88 & 63 & 0.00008 \\
			\textbf{SPNA} & 99 & 96 & 0.01528 & 92 & 74 & 0.08268 \\ \bottomrule
	\end{tabular} }
\end{table*}

Table~\ref{table:retrained_adversarial} presents the classification accuracy of perturbed samples. As the results show, almost all test samples are misclassified by the classifier model with very high accuracy at over 86\%. %With the exception of the Boundary Attack, all attacks successfully craft perturbations that are misclassified with over 86\% confidence.
Another important metric for the adversarial learning is the Mean Absolute Distance (MAD) and the Mean Squared Distance (MSD) of the perturbed traces from the originals. These metrics quantify the size of the changes i.e., perturbations made to the original traces by various adversarial attack methods. The difference between the MAD and the MSD is that, the latter is more sensitive to the larger changes due to the square operation. For instance, if an adversarial perturbation requires a significant change in 1 sample point among the 5000 features, it will have a stronger impact in the final MSD value than average change distributed over few points. MAD however is more dependent on the overall change in the trace, i.e., all 5000 sample points have the same impact on the final distance. Our results show that with most adversarial attacks, perturbation MAD is around or well below 1\% and within the ideal range. %Remember, the smaller the perturbation, easier it is to actuate it as a cloaking process.

\begin{table*}[th]
	\centering
	\caption{Effectiveness of adversarial re-training and DD on 100,000 perturbations crafted against the unprotected model. The values are the percentage of adversarial samples that are ineffective against models hardened by re-training and DD. While both methods manage to invalidate a small portion of the adversarial samples, most samples can still successfully fool the classifier.}
	\label{table:valid_pert}
	\resizebox{0.8\textwidth}{!}{%
		\begin{tabular}{@{}lcccccccccc@{}}
			\toprule
			\textbf{\begin{tabular}[c]{@{}l@{}}Adversarial\\   Attack\end{tabular}} & \textbf{\begin{tabular}[c]{@{}c@{}}Adversarial\\ Re-training\end{tabular}} & \textbf{\begin{tabular}[c]{@{}c@{}}DD with\\ T=1\end{tabular}} & \textbf{\begin{tabular}[c]{@{}c@{}}DD with\\ T=2\end{tabular}} & \textbf{\begin{tabular}[c]{@{}c@{}}DD with\\ T=5\end{tabular}} & \textbf{\begin{tabular}[c]{@{}c@{}}DD with\\ T=10\end{tabular}} & \textbf{\begin{tabular}[c]{@{}c@{}}DD with\\ T=20\end{tabular}} & \textbf{\begin{tabular}[c]{@{}c@{}}DD with\\ T=30\end{tabular}} & \textbf{\begin{tabular}[c]{@{}c@{}}DD with\\ T=40\end{tabular}} & \textbf{\begin{tabular}[c]{@{}c@{}}DD with\\ T=50\end{tabular}} & \textbf{\begin{tabular}[c]{@{}c@{}}DD with\\ T=100\end{tabular}} \\ \midrule
			\textbf{AGNA} & 42 & 77 & 60 & 70 & 83 & 83 & 63 & 64 & 62 & 75 \\
			\textbf{AUNA} & 43 & 77 & 60 & 70 & 83 & 82 & 63 & 65 & 61 & 75 \\
			\textbf{BUNA} & 94 & 92 & 92 & 91 & 94 & 94 & 94 & 96 & 94 & 100 \\
			\textbf{CRA} & 94 & 95 & 99 & 94 & 94 & 94 & 94 & 99 & 88 & 95 \\
			\textbf{GA} & 97 & 99 & 72 & 83 & 99 & 99 & 96 & 80 & 90 & 90 \\
			\textbf{GBA} & 84 & 83 & 84 & 88 & 82 & 94 & 93 & 91 & 93 & 88 \\
			\textbf{GSA} & 99 & 91 & 90 & 91 & 99 & 99 & 99 & 95 & 99 & 98 \\
			\textbf{LBFGSA} & 51 & 76 & 63 & 63 & 78 & 87 & 65 & 65 & 63 & 71 \\
			%			\textbf{SLSQPA} & 94.73532 & 94.72741 & 94.72741 & 94.73532 & 94.77222 & 100 & 94.72741 & 94.73004 & 94.72741 & 100 \\
			\textbf{SMA} & 26 & 71 & 50 & 52 & 62 & 82 & 49 & 47 & 32 & 48 \\
			\textbf{SPNA} & 4 & 84 & 76 & 78 & 94 & 93 & 76 & 79 & 73 & 80 \\ \bottomrule
		\end{tabular}%
	}	
\end{table*} %\vspace{25pt}

%\vspace{15pt}
\begin{table*}[th]
	\centering
	\caption{Perturbation MAD sizes of adversarial samples crafted against the unprotected, re-trained and DD hardened classifiers. Application of the adversarial re-training and DD only marginally increases the perturbation size needed to fool the classifier in most cases. By slightly increasing the perturbation size, all models can be fooled regardless of the applied defense method.}
	\label{table:dd_adversarial}
	\resizebox{0.8\textwidth}{!}{%
		\begin{tabular}{@{}lllcllllllll@{}}
			\toprule
			\textbf{\begin{tabular}[c]{@{}l@{}}Adversarial\\ Attack\end{tabular}} & \multicolumn{1}{c}{\textbf{\begin{tabular}[c]{@{}c@{}}Unprotected\\ Model\end{tabular}}} & \textbf{\begin{tabular}[c]{@{}l@{}}Adversarial\\ Re-trained\end{tabular}} & \textbf{\begin{tabular}[c]{@{}c@{}}DD with\\ T=1\end{tabular}} & \multicolumn{1}{c}{\textbf{\begin{tabular}[c]{@{}c@{}}DD with\\ T=2\end{tabular}}} & \multicolumn{1}{c}{\textbf{\begin{tabular}[c]{@{}c@{}}DD with\\ T=5\end{tabular}}} & \multicolumn{1}{c}{\textbf{\begin{tabular}[c]{@{}c@{}}DD with\\ T=10\end{tabular}}} & \multicolumn{1}{c}{\textbf{\begin{tabular}[c]{@{}c@{}}DD with\\ T=20\end{tabular}}} & \multicolumn{1}{c}{\textbf{\begin{tabular}[c]{@{}c@{}}DD with\\ T=30\end{tabular}}} & \multicolumn{1}{c}{\textbf{\begin{tabular}[c]{@{}c@{}}DD with\\ T=40\end{tabular}}} & \multicolumn{1}{c}{\textbf{\begin{tabular}[c]{@{}c@{}}DD with\\ T=50\end{tabular}}} & \multicolumn{1}{c}{\textbf{\begin{tabular}[c]{@{}c@{}}DD with\\ T=100\end{tabular}}} \\ \midrule
			\textbf{AGNA} & 0.00294 & 0.00294 & 0.00035 & 0.00265 & 0.00062 & 0.01059 & 0.01389 & 0.00452 & 0.00766 & 0.01797 & 0.00431 \\
			\textbf{AUNA} & 0.00292 & 0.00332 & 0.00033 & 0.00238 & 0.00071 & 0.01114 & 0.01336 & 0.00514 & 0.00948 & 0.01787 & 0.00451 \\
			%			\textbf{BA} & 0.00022 & 0.00228 &  &  &  &  &  &  &  &  &  \\
			\textbf{BUNA} & 0.05000 & 0.05000 & 0.04989 & 0.05301 & 0.08198 & 0.10293 & 7.52282 & 0.05006 & 0.05002 & 0.16376 & 0.07748 \\
			\textbf{CRA} & 0.05254 & 0.04999 & 0.04999 & 0.10247 & 0.08548 & 0.10780 & NA & 0.04998 & 1.44158 & 0.27992 & 0.07448 \\
			\textbf{GA} & 0.00250 & 0.00398 & 0.00355 & 0.00283 & 0.10224 & 0.00783 & 0.00364 & 0.00860 & 2.75805 & 0.00849 & 0.00303 \\
			\textbf{GBA} & 0.00080 & 0.00071 & 0.00058 & 0.00092 & 0.00060 & 0.04992 & NA & 0.00169 & 0.00046 & 0.00062 & 0.00083 \\
			\textbf{GSA} & 0.00499 & 0.00596 & 0.00504 & 0.00534 & 0.13665 & 0.02482 & 0.00540 & 0.01064 & 0.04121 & 0.01670 & 0.00550 \\
			\textbf{LBFGSA} & 0.00025 & 0.00031 & 0.00533 & 0.00178 & 0.00024 & 0.00210 & 0.07670 & 0.00088 & 0.02872 & 0.00521 & 0.00920 \\
			%			\textbf{RA} & 0.00005 & 0.00316 &  &  &  &  &  &  &  &  &  \\
			%	\textbf{SLSQPA} & 49.66867 & 49.66902 & 49.66888 & 49.66861 & 49.66878 & 49.66789 &  & 49.66792 & 49.66950 & 49.66875 & 49.66877 \\
			\textbf{SMA} & 0.00001 & 0.00008 & 0.00003 & 0.00012 & 0.00007 & NA & NA & 0.00001 & 0.00001 & NA & NA \\
			\textbf{SPNA} & 0.01528 & 0.08268 & 0.01060 & 0.00940 & 0.00360 & 0.01804 & 0.02000 & 0.00260 & 0.02000 & 0.02000 & 0.01980 \\ \bottomrule
		\end{tabular}%
	}
\end{table*}
\subsection{AL on the Hardened Models}%: Adversarial Re-training}
%%%%%%%%%%%%%%%%%%%%%%%%%%%%%%%%%%%%%%%%%%%%%%%%%%%%%%%%%%

Here we present the results of the AL attacks on the hardened classifier models. As explained in Section~\ref{sec:methodology}, we wanted to test the robustness of our cloaking mechanism against classifiers hardened with Adversarial Re-training and DD. To recap the scenario, Alice the defender wants to cloak her process by adding perturbations to her execution trace so that eavesdropper Eve cannot correctly classify what Alice is running. Then Eve notices or predicts the use of adversarial perturbations on the data and hardens her classifier model against AL attacks using adversarial re-training and DD. 

In order to test the attack scenario on hardened models, we first craft 100,000 adversarial samples per adversarial attack against the unprotected classifier. Then we harden the classifier with the aforementioned defense methods and feed the adversarial samples. Here, we aim to measure the level of protection provided by the adversarial re-training and the DD methods. As presented in Table~\ref{table:valid_pert}, the application of both the adversarial re-training and the DD invalidates some portion of the previously crafted adversarial samples. For the adversarial re-training, the success rate varies between 99\% (GSA) and 4\% (SPNA). In other words, 99\% of the adversarial samples crafted using GSA against the unprotected model are invalid on the hardened model. As for the DD, we see similar rates of protection ranging from 61\% up to 100\% for old perturbations. Impressively, 100\% of the adversarial samples crafted using the BUNA are now ineffective against the model trained with DD at temperature T=100. In short, by using the adversarial re-training or the DD, Eve can indeed harden her classifier against AL. However, keep in mind that Alice can observe or predict this behavior and introduce new adversarial samples targeting the hardened models. %Below, we discuss the results of our experiments against such hardened models.

%
%\begin{table}[t!]
%	\centering
%		\caption{Perturbation results against the CNN classifier hardened with adversarial re-training. The new adversarial samples have up to 29\% lower misclassification confidence rate compared to the unprotected model. However, adversarial samples are still misclassified with quite high confidence values in the range of 63-98\%.}
%	\label{table:retrained_adversarial}
%	\resizebox{\columnwidth}{!}{%
%		\begin{tabular}{@{}lccc@{}}
%			\toprule
%			\textbf{\begin{tabular}[c]{@{}l@{}}Adversarial\\ Attack\end{tabular}} & \textbf{\begin{tabular}[c]{@{}c@{}}Original Sample\\ Classification\\Confidence\end{tabular}} & \textbf{\begin{tabular}[c]{@{}c@{}}Perturbed Sample\\ MisClassification\\Confidence\end{tabular}} & \textbf{\begin{tabular}[c]{@{}c@{}}Mean\\Absolute\\ Distance\end{tabular}} \\ \midrule
%			\textbf{AGNA} & 92 & 82 & 0.00294 \\
%			\textbf{AUNA} & 91 &  82 & 0.00332 \\
%			%			\textbf{BA} & 92.99223 & 49.55626 & 0.00228 \\
%			\textbf{BUNA} & 96 & 93 & 0.05000 \\
%			\textbf{CRA} & 97 & 98 & 0.04999 \\
%			\textbf{GA} & 88 & 97 & 0.00398 \\
%			\textbf{GBA} & 93 & 74 & 0.00071 \\
%			\textbf{GSA} & 89 & 97 & 0.00596 \\
%			\textbf{LBFGSA} & 89 & 72 & 0.00031 \\
%			%			\textbf{RA} & 95.41001 & 48.45639 & 0.00316 \\
%			%		\textbf{SLSQPA} & 92.43184 & 100 & 49.66902 \\
%			\textbf{SMA} & 88 & 63 & 0.00008 \\
%			\textbf{SPNA} & 92 & 74 & 0.08268 \\ \bottomrule
%	\end{tabular} }
%\end{table}

\subsubsection{\textbf{Adversarial Re-training:}} After training the classifier model and crafting adversarial samples, we use these perturbations as training data and re-train the classifier. The motivation here is to teach the classifier model to detect these perturbed samples and correctly classify them. With this re-training stage, we expect to see whether we can `immunize' the model against given adversarial attacks. However, as the results in Table~\ref{table:retrained_adversarial} show, all of the adversarial attacks still succeed albeit requiring marginally larger perturbations. Moreover, while we observe a drop in the misclassification confidence, it is still quite high at over 63\% i.e., Eve's classifier can still be fooled by the adversarial samples.

\subsubsection{\textbf{Defensive Distillation:}} We have used the technique proposed in~\cite{papernot2016distillation} and trained hardened models with DD at various temperatures ranging from 1 to 100. Our results show that, even if the eavesdropper Eve hardens her model with DD, the trained model is still vulnerable to adversarial attacks albeit requiring larger perturbations in some cases. In Table~\ref{table:dd_adversarial}, we present the MAD i.e., the perturbation size, of various attack methods on both unprotected and hardened models. Our results show that the application of DD indeed offers a certain level of hardening to the model and increases the perturbation sizes. However this behavior is erratic compared to the adversarial re-training defense i.e., the MAD is significantly higher at some temperatures while much smaller for others. For instance, the MAD for the AUNA perturbations against the unprotected model is 0.00292 in average for 100,000 adversarial samples. The perturbation size for the same attack drops to 0.00033 when the distillation defense is applied with temperature T=1. This in turn practically makes Eve's classifier model easier to fool. Same behavior is observed with the adversarial samples crafted using AGNA and GBA as well. For the cases that the DD actually hardens Eve's model against adversarial samples, the MAD is still minimal. Hence, Alice can still successfully craft adversarial samples with minimal perturbations and fool Eve's model. Finally, NA values in Table~\ref{table:dd_adversarial} represent cases where the model had very low classification accuracy and could not correctly classify original samples.

%~\todo[inline]{discuss more; indeed, the results shold be more closely mapped to the new figure 1. Right now the contribution is not obvious and I get lost in who is the adversary in shich scenario.}%Moreover, even when the previously crafted adversarial samples are fed into the distilled networks, the successful misclassification i.e., the successful cloaking rate 

%\subsubsection{\textbf{Some Empirical Observations:}}
%
%We have crafted adversarial perturbations on models that are trained with varying number of training samples ranging from 100 samples per class up to 30,000. After perturbation creation, we have observed that the perturbations on highly trained models have larger MAD and MSD values. This shows that it is actually harder to fool well-trained models than poorly-trained models. Moreover, when only two HPC traces (Total Instructions and Branch Instructions) are used, the classification rate is quite high at 99.5\%.
%%%%%%%%%%%%%%%%%%%%%%%%%
\section{Conclusion}\label{sec:conclusion}
%%%%%%%%%%%%%%%%%%%%%%%%%

Side-channel leakage on shared hardware systems pose a real and present danger to the security and the privacy of users. Even when the software is perfectly isolated, co-resident tenants still share the underlying hardware and are prone to side-channel attacks. Especially considering the wide adoption of AI across many disciplines, it is not surprising that such attacks will become automated and even easier to perform in the future. There is a clear need for users to cloak their execution fingerprints from the underlying shared system.

With this work we took a first step in this direction. Specifically, by making clever defensive use of adversarial crafting we introduced a new cloaking defense against the side-channel leakage classifiers. We first demonstrated the threat side-channel leakage poses by processing leakage profiles to yield highly accurate AI models which may be used by an adversary to violate privacy and security policies of applications. We trained various types of classifiers including the classical ML methods and showed how the parameter selection affects the learning rate and the validation accuracy. While this is a strong threat to shared hardware systems, we showed that it can be mitigated using carefully crafted adversarial samples. Moreover, we investigated defenses that can potentially help an attack to bypass the adversarial samples. Our results show that even in the presence of defensive distillation and adversarial re-training, the defender can craft working adversarial samples and fool the attacker. These perturbations can be implemented as a sister-process that will run side-by-side with the original and easily cause misclassification to the attacker's model without any significant overhead. Using the adversarial crafting-based cloaking mechanism that we have outlined in this work, users can enable such services on-demand for sensitive operations. Efficient design and implementation of such defenses for shared hardware systems like cloud remains an open research problem. Finally, to the best of our knowledge, this work is the first use of adversarial crafting for defensive purposes. We envision the same approach to be useful in other application scenarios.

%A malicious adversary can obtain detailed information about the processes running on a shared hardware by obtaining the HPC trace of a process or the overall system. On the bright side, this automation comes with its inherent weakness against adversarial learning and this weakness can be exploited by the defender. We showed that by using variety of adversarial learning methods, we can craft very small perturbations to cloak a process and defend against attackers equipped with AI tools. These perturbations can be implemented as a sister-process that will run side-by-side with the original process and easily cause misclassification to the attacker's model without any significant overhead.

%Finally, we believe that the malicious actors will add more and more AI and ML techniques to their arsenal over time. To counter this new attack vector, tailored countermeasures should be designed and put in place ahead of time by exploiting inherent weaknesses in AI systems.

%\clearpage
\bibliographystyle{acm}\balance
\bibliography{mybib}

\begin{thebibliography}{10}

\bibitem{alamperformance}
{\sc Alam, M., Bhattacharya, S., Mukhopadhyay, D., and Bhattacharya, S.}
\newblock Performance counters to rescue: A machine learning based safeguard
  against micro-architectural side-channel-attacks.
\newblock Cryptology ePrint Archive, Report 2017/564, 2017.
\newblock \url{https://eprint.iacr.org/2017/564}.

\bibitem{ML_SCA_detection}
{\sc Allaf, Z., Adda, M., and Gegov, A.}
\newblock A comparison study on flush+reload and prime+probe attacks on aes
  using machine learning approaches.
\newblock In {\em Advances in Computational Intelligence Systems\/} (Cham,
  2018), F.~Chao, S.~Schockaert, and Q.~Zhang, Eds., Springer International
  Publishing, pp.~203--213.

\bibitem{deep_lip_reading_paper}
{\sc Assael, Y.~M., Shillingford, B., Whiteson, S., and de~Freitas, N.}
\newblock Lipnet: Sentence-level lipreading.
\newblock {\em arXiv preprint arXiv:1611.01599\/} (2016).

\bibitem{beltramelli2015deep}
{\sc Beltramelli, T., and Risi, S.}
\newblock Deep-spying: Spying using smartwatch and deep learning.
\newblock {\em arXiv preprint arXiv:1512.05616\/} (2015).

\bibitem{Bernstein05cache-timingattacks}
{\sc Bernstein, D.~J.}
\newblock {Cache-timing attacks on AES}, 2004.
\newblock {URL}: {\tt http://cr.yp.to/papers.html\#cachetiming}.

\bibitem{biggio2013evasion}
{\sc Biggio, B., Corona, I., Maiorca, D., Nelson, B., {\v{S}}rndi{\'c}, N.,
  Laskov, P., Giacinto, G., and Roli, F.}
\newblock Evasion attacks against machine learning at test time.
\newblock In {\em Joint European conference on machine learning and knowledge
  discovery in databases\/} (2013), Springer, pp.~387--402.

\bibitem{biggio2014security}
{\sc Biggio, B., Fumera, G., and Roli, F.}
\newblock Security evaluation of pattern classifiers under attack.
\newblock {\em IEEE transactions on knowledge and data engineering 26}, 4
  (2014), 984--996.

\bibitem{biggio2011support}
{\sc Biggio, B., Nelson, B., and Laskov, P.}
\newblock Support vector machines under adversarial label noise.
\newblock In {\em Asian Conference on Machine Learning\/} (2011), pp.~97--112.

\bibitem{biggio2012poisoning}
{\sc Biggio, B., Nelson, B., and Laskov, P.}
\newblock Poisoning attacks against support vector machines.
\newblock {\em arXiv preprint arXiv:1206.6389\/} (2012).

\bibitem{samira_cacheshield_sca_protection}
{\sc Briongos, S., Irazoqui, G., Malag{\'o}n, P., and Eisenbarth, T.}
\newblock Cacheshield: Protecting legacy processes against cache attacks.
\newblock {\em arXiv preprint arXiv:1709.01795\/} (2017).

\bibitem{samira_cacheshield2_sca_detection}
{\sc Briongos, S., Irazoqui, G., Malag\'{o}n, P., and Eisenbarth, T.}
\newblock Cacheshield: Detecting cache attacks through self-observation.
\newblock In {\em Proceedings of the Eighth ACM Conference on Data and
  Application Security and Privacy\/} (New York, NY, USA, 2018), CODASPY '18,
  ACM, pp.~224--235.

\bibitem{carlini2017towards}
{\sc Carlini, N., and Wagner, D.}
\newblock Towards evaluating the robustness of neural networks.
\newblock In {\em Security and Privacy (SP), 2017 IEEE Symposium on\/} (2017),
  IEEE, pp.~39--57.

\bibitem{quickhpc}
{\sc Chiappetta, M.}
\newblock Quickhpc.
\newblock \url{https://github.com/chpmrc/quickhpc}, 2015.

\bibitem{dang2017evading}
{\sc Dang, H., Huang, Y., and Chang, E.-C.}
\newblock Evading classifiers by morphing in the dark.
\newblock In {\em Proceedings of the 2017 ACM SIGSAC Conference on Computer and
  Communications Security\/} (2017), ACM, pp.~119--133.

\bibitem{AIphishing}
{\sc Emmanuel, Z.}
\newblock Security experts air concerns over hackers using ai and machine
  learning for phishing attacks.
\newblock
  \url{https://www.computerweekly.com/news/450427653/Security-experts-air-concerns-over-hackers-using-AI-and
  -machine-learning-for-phishing-atttacks/}.

\bibitem{eykholt2017note}
{\sc Eykholt, K., Evtimov, I., Fernandes, E., Li, B., Song, D., Kohno, T.,
  Rahmati, A., Prakash, A., and Tramer, F.}
\newblock Note on attacking object detectors with adversarial stickers.
\newblock {\em arXiv preprint arXiv:1712.08062\/} (2017).

\bibitem{foreman2018survey}
{\sc Foreman, J.~C.}
\newblock A survey of cyber security countermeasures using hardware performance
  counters.
\newblock {\em arXiv preprint arXiv:1807.10868\/} (2018).

\bibitem{attacking_ml}
{\sc Goodfellow, I., and Papernot, N.}
\newblock {Is attacking machine learning easier than defending it?}
\newblock
  \url{http://www.cleverhans.io/security/privacy/ml/2017/02/15/why-attacking-machine-learning-is-easier-than-defending-it.html}.

\bibitem{goodfellow2014explaining}
{\sc Goodfellow, I.~J., Shlens, J., and Szegedy, C.}
\newblock Explaining and harnessing adversarial examples.
\newblock {\em arXiv preprint arXiv:1412.6572\/} (2014).

\bibitem{deep_smash2}
{\sc Gordon, R.}
\newblock {AI beats pros at Super Smash Bros.}
\newblock \url{http://www.csail.mit.edu/ai_beats_pros_at_super_smash_bros}.
\newblock Accessed: 2017-10-27.

\bibitem{gruss2016flush+}
{\sc Gruss, D., Maurice, C., Wagner, K., and Mangard, S.}
\newblock Flush+ flush: a fast and stealthy cache attack.
\newblock In {\em International Conference on Detection of Intrusions and
  Malware, and Vulnerability Assessment\/} (2016), Springer, pp.~279--299.

\bibitem{gruss2015cache}
{\sc Gruss, D., Spreitzer, R., and Mangard, S.}
\newblock Cache template attacks: automating attacks on inclusive last-level
  caches.
\newblock In {\em Proceedings of the 24th USENIX Conference on Security
  Symposium\/} (2015), USENIX Association, pp.~897--912.

\bibitem{gullasch2011cache}
{\sc Gullasch, D., Bangerter, E., and Krenn, S.}
\newblock {Cache games--bringing access-based cache attacks on AES to
  practice}.
\newblock In {\em Security and Privacy (SP), 2011 IEEE Symposium on\/} (2011),
  IEEE.

\bibitem{gulmezoglu2017perfweb}
{\sc G{\"{u}}lmezoglu, B., Zankl, A., Eisenbarth, T., and Sunar, B.}
\newblock {PerfWeb: How to Violate Web Privacy with Hardware Performance
  Events}.
\newblock In {\em Computer Security - {ESORICS} 2017 - 22nd European Symposium
  on Research in Computer Security, Oslo, Norway, September 11-15, 2017,
  Proceedings, Part {II}\/} (2017), pp.~80--97.

\bibitem{deep_diabetic_eye_paper}
{\sc Gulshan, V., Peng, L., Coram, M., Stumpe, M.~C., Wu, D., Narayanaswamy,
  A., Venugopalan, S., Widner, K., Madams, T., Cuadros, J., et~al.}
\newblock Development and validation of a deep learning algorithm for detection
  of diabetic retinopathy in retinal fundus photographs.
\newblock {\em Jama 316}, 22 (2016).

\bibitem{deep_feats}
{\sc Hadad, Y.}
\newblock 30 amazing applications of deep learning.
\newblock
  \url{http://www.yaronhadad.com/deep-learning-most-amazing-applications/}, Aug
  2017.
\newblock Accessed: 2017-10-27.

\bibitem{heuser2012intelligent}
{\sc Heuser, A., and Zohner, M.}
\newblock Intelligent machine homicide.
\newblock {\em COSADE 7275\/} (2012), 249--264.

\bibitem{hospodar2011machine}
{\sc Hospodar, G., Gierlichs, B., De~Mulder, E., Verbauwhede, I., and
  Vandewalle, J.}
\newblock Machine learning in side-channel analysis: a first study.
\newblock {\em Journal of Cryptographic Engineering 1}, 4 (2011), 293.

\bibitem{huang2011adversarial}
{\sc Huang, L., Joseph, A.~D., Nelson, B., Rubinstein, B.~I., and Tygar, J.}
\newblock Adversarial machine learning.
\newblock In {\em Proceedings of the 4th ACM workshop on Security and
  artificial intelligence\/} (2011), ACM, pp.~43--58.

\bibitem{inci2016cache}
{\sc Inci, M.~S., Gulmezoglu, B., Irazoqui, G., Eisenbarth, T., and Sunar, B.}
\newblock Cache attacks enable bulk key recovery on the cloud.
\newblock In {\em International Conference on Cryptographic Hardware and
  Embedded Systems---CHES\/} (2016), Springer Berlin Heidelberg, pp.~368--388.

\bibitem{waitaminute}
{\sc Irazoqui, G., Inci, M.~S., Eisenbarth, T., and Sunar, B.}
\newblock Wait a minute! a fast, cross-vm attack on aes.
\newblock In {\em International Workshop on Recent Advances in Intrusion
  Detection\/} (2014), Springer, pp.~299--319.

\bibitem{fcc_ars}
{\sc {Jon Brodkin}}.
\newblock 2 million people-and some dead ones-were impersonated in net
  neutrality comments.
\newblock
  \url{https://arstechnica.com/tech-policy/2017/12/dead-people-among-millions-impersonated-in-fake-
  net-neutrality-comments/}.

\bibitem{ai_hacking}
{\sc {Keren Elazari}}.
\newblock {Hackers are on the brink of launching a wave of AI attacks}.
\newblock
  \url{http://www.wired.co.uk/article/hackers-ai-cyberattack-offensive}.

\bibitem{kurakin2016adversarial}
{\sc Kurakin, A., Goodfellow, I., and Bengio, S.}
\newblock Adversarial machine learning at scale.
\newblock {\em arXiv preprint arXiv:1611.01236\/} (2016).

\bibitem{laskov2010machine}
{\sc Laskov, P., and Lippmann, R.}
\newblock Machine learning in adversarial environments, 2010.

\bibitem{lipp2016armageddon}
{\sc Lipp, M., Gruss, D., Spreitzer, R., Maurice, C., and Mangard, S.}
\newblock Armageddon: Cache attacks on mobile devices.
\newblock In {\em 25th {USENIX} Security Symposium ({USENIX} Security 16)\/}
  (2016), {USENIX} Association, pp.~235--252.

\bibitem{liu2016delving}
{\sc Liu, Y., Chen, X., Liu, C., and Song, D.}
\newblock Delving into transferable adversarial examples and black-box attacks.
\newblock {\em arXiv preprint arXiv:1611.02770\/} (2016).

\bibitem{deep_breast_cancer_paper}
{\sc Liu, Y., Gadepalli, K., Norouzi, M., Dahl, G.~E., Kohlberger, T., Boyko,
  A., Venugopalan, S., Timofeev, A., Nelson, P.~Q., Corrado, G.~S., et~al.}
\newblock Detecting cancer metastases on gigapixel pathology images.
\newblock {\em arXiv preprint arXiv:1703.02442\/} (2017).

\bibitem{lowd2005adversarial}
{\sc Lowd, D., and Meek, C.}
\newblock Adversarial learning.
\newblock In {\em Proceedings of the eleventh ACM SIGKDD international
  conference on Knowledge discovery in data mining\/} (2005), ACM,
  pp.~641--647.

\bibitem{deep_iq2}
{\sc MacDonald, F.}
\newblock {A Deep Learning Machine Just Beat Humans in an IQ Test}.
\newblock
  \url{https://www.sciencealert.com/a-deep-learning-machine-just-beat-humans-in-an-iq-test}.

\bibitem{maghrebi2016breaking}
{\sc Maghrebi, H., Portigliatti, T., and Prouff, E.}
\newblock Breaking cryptographic implementations using deep learning
  techniques.
\newblock In {\em International Conference on Security, Privacy, and Applied
  Cryptography Engineering\/} (2016), Springer, pp.~3--26.

\bibitem{martinasek2013optimization}
{\sc Martinasek, Z., Hajny, J., and Malina, L.}
\newblock Optimization of power analysis using neural network.
\newblock In {\em International Conference on Smart Card Research and Advanced
  Applications\/} (2013), Springer, pp.~94--107.

\bibitem{martinasek2013innovative}
{\sc Martinasek, Z., and Zeman, V.}
\newblock Innovative method of the power analysis.
\newblock {\em Radioengineering 22}, 2 (2013), 586--594.

\bibitem{meng2017magnet}
{\sc Meng, D., and Chen, H.}
\newblock Magnet: a two-pronged defense against adversarial examples.
\newblock In {\em Proceedings of the 2017 ACM SIGSAC Conference on Computer and
  Communications Security\/} (2017), ACM, pp.~135--147.

\bibitem{deep_romance_novel}
{\sc O'Brien, E.}
\newblock Romance novels, generated by artificial intelligence.
\newblock
  \url{https://medium.com/towards-data-science/romance-novels-generated-by-artificial-intelligence-
  1b31d9c872b2}, Aug 2017.

\bibitem{oh2017adversarial}
{\sc Oh, S.~J., Fritz, M., and Schiele, B.}
\newblock Adversarial image perturbation for privacy protection a game theory
  perspective.
\newblock In {\em 2017 IEEE International Conference on Computer Vision
  (ICCV)\/} (2017), IEEE, pp.~1491--1500.

\bibitem{osvik2006cache}
{\sc Osvik, D.~A., Shamir, A., and Tromer, E.}
\newblock {Cache attacks and countermeasures: the case of AES}.
\newblock In {\em Cryptographers Track at the RSA Conference\/} (2006),
  Springer, pp.~1--20.

\bibitem{papernot2016transferability}
{\sc Papernot, N., McDaniel, P., and Goodfellow, I.}
\newblock Transferability in machine learning: from phenomena to black-box
  attacks using adversarial samples.
\newblock {\em arXiv preprint arXiv:1605.07277\/} (2016).

\bibitem{papernot2017practical}
{\sc Papernot, N., McDaniel, P., Goodfellow, I., Jha, S., Celik, Z.~B., and
  Swami, A.}
\newblock Practical black-box attacks against machine learning.
\newblock In {\em Proceedings of the 2017 ACM on Asia Conference on Computer
  and Communications Security\/} (2017), ACM, pp.~506--519.

\bibitem{papernot2016limitations}
{\sc Papernot, N., McDaniel, P., Jha, S., Fredrikson, M., Celik, Z.~B., and
  Swami, A.}
\newblock The limitations of deep learning in adversarial settings.
\newblock In {\em Security and Privacy (EuroS\&P), 2016 IEEE European Symposium
  on\/} (2016), IEEE, pp.~372--387.

\bibitem{papernot2016distillation}
{\sc Papernot, N., McDaniel, P., Wu, X., Jha, S., and Swami, A.}
\newblock Distillation as a defense to adversarial perturbations against deep
  neural networks.
\newblock In {\em Security and Privacy (SP), 2016 IEEE Symposium on\/} (2016),
  IEEE, pp.~582--597.

\bibitem{papernot_saliency}
{\sc Papernot, N., McDaniel, P.~D., Jha, S., Fredrikson, M., Celik, Z.~B., and
  Swami, A.}
\newblock The limitations of deep learning in adversarial settings.
\newblock {\em CoRR abs/1511.07528\/} (2015).

\bibitem{deep_starcraft_paper}
{\sc Peng, P., Yuan, Q., Wen, Y., Yang, Y., Tang, Z., Long, H., and Wang, J.}
\newblock {Multiagent Bidirectionally-Coordinated Nets for Learning to Play
  StarCraft Combat Games}.
\newblock {\em arXiv preprint arXiv:1703.10069\/} (2017).

\bibitem{percival2005cache}
{\sc Percival, C.}
\newblock Cache missing for fun and profit, 2005.

\bibitem{pessl2016drama}
{\sc Pessl, P., Gruss, D., Maurice, C., Schwarz, M., and Mangard, S.}
\newblock Drama: Exploiting dram addressing for cross-cpu attacks.
\newblock In {\em USENIX Security Symposium\/} (2016), pp.~565--581.

\bibitem{rauber2017foolbox}
{\sc Rauber, J., Brendel, W., and Bethge, M.}
\newblock Foolbox v0.8.0: A python toolbox to benchmark the robustness of
  machine learning models.
\newblock {\em arXiv preprint arXiv:1707.04131\/} (2017).

\bibitem{AIspam}
{\sc Simonite, T.}
\newblock {This AI Will Craft Tweets That Youll Never Know Are Spam}.
\newblock
  \url{https://www.technologyreview.com/s/602109/this-ai-will-craft-tweets-that-youll-never-know-are-spam/}.

\bibitem{su2017one}
{\sc Su, J., Vargas, D.~V., and Kouichi, S.}
\newblock One pixel attack for fooling deep neural networks.
\newblock {\em arXiv preprint arXiv:1710.08864\/} (2017).

\bibitem{fcc_bloomberg}
{\sc {Susan Decker}}.
\newblock Fcc rules out delaying net neutrality repeal over fake comments.
\newblock
  \url{https://www.bloomberg.com/news/articles/2018-01-05/fcc-rules-out-delaying-net-neutrality-repeal-over-
  fake-comments}.

\bibitem{szegedy2013intriguing}
{\sc Szegedy, C., Zaremba, W., Sutskever, I., Bruna, J., Erhan, D., Goodfellow,
  I., and Fergus, R.}
\newblock Intriguing properties of neural networks.
\newblock {\em arXiv preprint arXiv:1312.6199\/} (2013).

\bibitem{LBFGSA_paper}
{\sc Tabacof, P., and Valle, E.}
\newblock Exploring the space of adversarial images.
\newblock {\em CoRR abs/1510.05328\/} (2015).

\bibitem{ai_cyber}
{\sc {The Cylance Team}}.
\newblock {Black Hat Attendees See AI as Double-Edged Sword}.
\newblock
  \url{https://threatmatrix.cylance.com/en_us/home/black-hat-attendees-see-ai-as-double-edged-sword.html}.

\bibitem{tramer2017ensemble}
{\sc Tram{\`e}r, F., Kurakin, A., Papernot, N., Boneh, D., and McDaniel, P.}
\newblock Ensemble adversarial training: Attacks and defenses.
\newblock {\em arXiv preprint arXiv:1705.07204\/} (2017).

\bibitem{deep_speech_paper}
{\sc Xiong, W., Droppo, J., Huang, X., Seide, F., Seltzer, M., Stolcke, A., Yu,
  D., and Zweig, G.}
\newblock Achieving human parity in conversational speech recognition.
\newblock {\em arXiv preprint arXiv:1610.05256\/} (2016).

\bibitem{yaromFnR}
{\sc Yarom, Y., and Falkner, K.}
\newblock {FLUSH+RELOAD: A High Resolution, Low Noise, L3 Cache Side-Channel
  Attack}.
\newblock In {\em (USENIX Security 2014)\/} (2014).

\bibitem{deep_lung_cancer_paper}
{\sc Yu, K.-H., Zhang, C., Berry, G.~J., Altman, R.~B., R{\'e}, C., Rubin,
  D.~L., and Snyder, M.}
\newblock Predicting non-small cell lung cancer prognosis by fully automated
  microscopic pathology image features.
\newblock {\em Nature communications 7\/} (2016).

\bibitem{zhang2016cloudradar}
{\sc Zhang, T., Zhang, Y., and Lee, R.~B.}
\newblock {Cloudradar: A real-time side-channel attack detection system in
  clouds}.
\newblock In {\em International Symposium on Research in Attacks, Intrusions,
  and Defenses\/} (2016), Springer.

\bibitem{zhang2019statistical}
{\sc Zhang, X., Hamm, J., Reiter, M.~K., and Zhang, Y.}
\newblock Statistical privacy for streaming traffic.
\newblock In {\em NDSS\/} (2019).

\bibitem{zhang2014cross}
{\sc Zhang, Y., Juels, A., Reiter, M.~K., and Ristenpart, T.}
\newblock {Cross-tenant side-channel attacks in {PaaS} clouds}.
\newblock In {\em CCS\/} (2014), pp.~990--1003.

\bibitem{zhou2012adversarial}
{\sc Zhou, Y., Kantarcioglu, M., Thuraisingham, B., and Xi, B.}
\newblock Adversarial support vector machine learning.
\newblock In {\em Proceedings of the 18th ACM SIGKDD international conference
  on Knowledge discovery and data mining\/} (2012), ACM, pp.~1059--1067.

\end{thebibliography}
%\clearpage
%\input{appendix}
\end{document}